\documentclass{aa}
\usepackage{graphicx}
\usepackage[utf8]{inputenc}
\usepackage{txfonts}
\usepackage{times}
\usepackage{longtable}
\usepackage{natbib}
\usepackage{wasysym}
\usepackage{pdflscape}
\usepackage{caption}
\usepackage{subcaption}
\usepackage{placeins}
\usepackage{xcolor}
\usepackage[switch]{lineno}
\usepackage{tablefootnote}
\usepackage{hyperref}
\usepackage{ulem}
\hypersetup{
    colorlinks=true,
    linkcolor=magenta,
    citecolor=blue,
    filecolor=magenta,      
    urlcolor=cyan,
}
\usepackage{siunitx}
\usepackage{threeparttable}
\usepackage{dblfloatfix}    % To enable figures at the bottom of page
\usepackage{breqn}
\usepackage{makecell}

\DeclareMathOperator{\sech}{sech}

\DeclareSIUnit\gauss{G}
\DeclareSIUnit\erg{erg}
\DeclareSIUnit\pc{pc}
\DeclareSIUnit\year{yrs}

\begin{document}

\title{Galactic gamma-ray and neutrino emission from interacting cosmic-ray nuclei}
\titlerunning{Galactic $\gamma$-ray and neutrino emission from interacting CR nuclei}

\author{M. Breuhaus \inst{1,\star}, 
J. A. Hinton \inst{1}, 
V. Joshi \inst{2}, B. Reville \inst{1}, H. Schoorlemmer\inst{1,3,4} }

\offprints{corresponding author:  Mischa Breuhaus, \email{mischa.breuhaus@mpi-hd.mpg.de}\\Member of the International Max-Planck Research School for Astronomy and Cosmic Physics at the University of Heidelberg (IMPRS-HD), Germany}

\authorrunning{Breuhaus et al.}
\institute{Max-Planck-Institut f\"ur Kernphysik, Postfach 103980, D 69029 Heidelberg, Germany 
\and
Friedrich-Alexander-Universit\"{a}t Erlangen-N\"{u}rnberg, ECAP, Erwin-Rommel-Str. 1, D 91058 Erlangen, Germany
\and
 IMAPP, Radboud University Nijmegen, Nijmegen, The Netherlands
 \and
 Nationaal Instituut voor Kernfysica en Hoge Energie Fysica (NIKHEF), Science Park, Amsterdam, The Netherlands}  
\date{Accepted 2022-01-07; Received 2021-05-14}

\abstract{
We present a study of the expectations for very-high-energy (VHE) to ultra-high-energy (UHE) gamma-ray and neutrino emission from interacting cosmic rays in our Galaxy as well as a comparison to the latest results for the Galactic UHE diffuse emission. We demonstrate the importance of properly accounting for both the mixed cosmic-ray composition and the gamma-ray absorption. 
We adopt the wounded-nucleon model of nucleus interactions and provide parameterisations of the resulting gamma-ray and neutrino production.
Nucleon shielding due to clustering inside nuclei is shown to have a measurable effect on the production of gamma rays and is particularly evident close to breaks and cutoffs in mixed-composition particle spectra.
The change in composition around the `knee' in the cosmic ray spectrum has a noticeable impact on the diffuse neutrino and gamma-ray emission spectra. We show that current and near-future detectors can probe these differences in the key energy range from \SI{10}{\tera\electronvolt} to \SI{1}{\peta\electronvolt}, testing the paradigm of the universality of the cosmic ray spectrum and composition throughout the Galaxy. 
}


\maketitle

\section{Introduction}
\label{sec:intro}

Collisions between energetic cosmic rays (CRs) and target nuclei in astrophysical environments generate neutral ($\pi^0$) and charged pions ($\pi^{\pm}$), which decay to produce $\gamma$ rays, electrons, positrons, and neutrinos. The secondary products, which encode details of both the CR and target populations, are in principle measurable at Earth. The accurate interpretation of such measurements is necessary to understand both the acceleration and propagation of CRs in our Galaxy. The charge and mass of both beam and target species have an impact on secondary production \cite[see][and references therein]{Kafexhiu_et_al_2014}. Yet this effect is often ignored in high-energy astrophysics,
where hadronic interactions are commonly restricted to the idealised case of CR protons colliding with hydrogen nuclei.

At energies close to the threshold energy for pion production and below, effects such as the sub-threshold pion production or hard photon emission occur, which have previously been investigated in detail \citep{Kafexhiu2016, Yang_et_al_2018}. At higher energies, in many cases the contribution of nuclei can be neglected or captured in a modest scaling factor \cite[see for example][]{Mori_2009}. However, around spectral breaks or cutoffs in the particle spectra, approximating the influence of nuclei in the CR beam via simple energy-independent scaling factors is known to give erroneous predictions \cite[e.g.][]{Kachelriess_et_al_2014}. Thus, when modelling hadronic $\gamma$-ray and neutrino emission from sources, which for most cases will have a maximum particle kinetic energy that is rigidity limited, consideration of the composition is necessary. 

The same holds true for modelling diffuse $\gamma$-ray and neutrino emission from Galactic CR collisions with the interstellar medium (ISM) in the Milky Way at energies around the `knee' feature, at a few \si{\peta\electronvolt} in the all-particle CR spectrum.
Improvements in $\gamma$-ray \citep{Tibet_2019,Tibet_2021, HAWC_2020, LHAASO_2016, LHAASO_2019,CTA_KeyScience,SWGO_2019, SWGO_2019b} and neutrino \citep{IceCubeGen2_2017, IceCubeGen2_2020} detection capabilities in the \si{\tera\electronvolt} to \si{\peta\electronvolt} energy range enable multi-messenger probing in the window specifically relevant to the CR knee.
This energy range has a particular significance, as the knee feature in the all-particle CR spectrum plays a central role in the theory of Galactic CR origins
\cite[e.g.][]{HillasKnee}. 
The physical mechanism underlying the formation of the knee is an open question. Although supernova remnants remain the most plausible candidate for most Galactic CR production, the current theoretical understanding indicates that favourable conditions are already necessary to accelerate \si{\peta\electronvolt} protons, conditions that likely only occur in a subset of supernova remnants \citep{Belletal13,Marcowithetal18}. 

Observations that indicate \si{\peta\electronvolt} CR sources associated with star forming regions are emerging \cite[see for example][]{AharonianSFRs,BykovReview}, as are candidate sources in the Galactic centre region \citep{HESS_GCR}. Convincing evidence for the mechanism(s) that can account for the \si{\peta\electronvolt} CR flux at Earth, let alone the break in the all-particle spectrum at %\SIrange[range-phrase=-,range-units=single]{\sim 3}{4}{\peta\electronvolt}, 
$\sim 3-4~\si{\peta\electronvolt}$ is, however, still lacking.  
With the next generation of $\gamma$-ray instruments, the identification of Galactic sources of \si{\peta\electronvolt} CRs is anticipated \citep{CTA_KeyScience}. At the same time, sensitivity to the diffuse Galactic $\gamma$-ray and neutrino emission will see major improvements in the near future. In both cases, the ability to accurately interpret and distinguish spectral features, including the possible influence of composition effects, is paramount and can provide validation of existing and future theories.
We note, as an example, that a significant deviation in the expected Galactic diffuse emission associated with the knee feature would necessitate a revision of our current paradigm of CR origins.  

Owing to its importance for the interpretation of the emission, both in very-high-energy (VHE) $\gamma$-ray sources and the diffuse emission in the Galaxy, we consider in the following the impact of target and beam composition on the resulting hadronic emission and its signatures. To address these problems, the $\gamma$-ray and neutrino emission for CRs and target species of arbitrary composition and input spectrum have been incorporated into the open-source GAMERA code \citep{gamera}. The results are applied to two relevant cases: rigidity-limited accelerators and diffuse Galactic emission.

The Tibet AS$\gamma$ Collaboration recently reported on the first detection of diffuse $\gamma$-ray emission from the Galactic disk above \SI{100}{\tera\electronvolt} \citep{Tibet_2021}. 
While the diffuse ultra-high-energy (UHE) $\gamma$-ray and neutrino emission from the Galaxy has been previously considered \citep{Lipari_Vernetto_2018}, details of the composition had not been directly addressed. In the following, we focus specifically on the composition effects close to the knee in light of these new observational results, comparing $\gamma$-ray and neutrino expectations, and identify signatures that might be measurable in the near future.  

The outline of the paper is as follows. In Sect. \ref{sec:methodology} we first describe the method used to account for the composition in the production of $\gamma$ rays, neutrinos, and electrons and positrons in hadronic collisions. This is followed by an investigation of the effects relevant for sources with a rigidity-dependent cutoff and the development of a simple approximation for the $\gamma$-ray and neutrino emission from exponential cutoff power-law distributed CRs. In Sect. \ref{sec:diffuse_emission} we turn to the investigation of the implications for the diffuse $\gamma$-ray and neutrino emission from the Galactic plane. We conclude with a summary and discussion of our results (Sect. \ref{sec:conclusion}). Additional details are provided in the appendix.

\section{Modelling $\gamma$-ray and neutrino production in sources with rigidity-limited cutoff} 
\label{sec:methodology}

For all conceivable astrophysical sources of CRs, the maximum energy is rigidity dependent. The composition of the accelerated particle distribution depends on the environmental conditions in the vicinity of the accelerator. For example, most young supernova shocks propagate into the winds of their progenitors, which may have a  significantly depleted hydrogen content relative to the typical ISM \cite[e.g.][]{Langer}. If \si{\peta\electronvolt} CRs are primarily produced in the very early stages of core-collapse supernovae \cite[e.g.][]{Belletal13}, understanding the hadronic radiative signatures of these sources requires an accurate picture of composition effects beyond the idealised scenario of proton-proton collision.

The $\gamma$-ray and neutrino production from nucleus-nucleus collisions was implemented in the open-source GAMERA code \citep{gamera}. The algorithm for producing the spectrum followed closely the method developed in \cite{Kelner2006} updated to include improvements discussed in \citet{Kafexhiu_et_al_2014}. 
Collisions between nuclei in this work are approximated with the so-called wounded nucleon model \citep{Biallas1976,Rybczynski2011}, in which the collision is described as a series of interactions between individual nucleons. In this approximation, effects that may result from the motion or excitation inside the whole nucleus are neglected. Production of secondary particles, such as neutral or charged pions, is the sum of all individual nucleon-nucleon secondary particle productions. The collision between individual nucleons is thus treated as a collision between two protons.
The total production rates for the different products are in general given by 
\begin{align}
&\frac{\text dN(E)}{\text dE\,\,\text dt} =\\ 
&~~~~\sum_{i,j} c n_j \int_{E}^{\infty} \kappa(E_{\text p,i}) \sigma_{\rm inel}(E_{\text p,i}) J_{\text p,i}(E_{\text p,i}) F\left( \frac{E}{E_{\text p,i}}, E_{\text p,i}\right) \frac{\text d E_{\text p,i}}{E_{\text p,i}}~.
   \nonumber
\end{align}
Here, the index $i$ refers to the respective CR species and the index $j$ to the target species. $E_{\text p,i} = E_{\text p}/A_i$ is the energy per nucleon of CR species $i$ with mass number $A_i$, $\sigma_{\rm inel}$ the inelastic cross-section from \citet{Kafexhiu_et_al_2014}, $J_{\text p,i}$ the energy number density of CR species $i$, and $F(x, E_p)$ is the differential number of photons/neutrinos produced between $x$ and $x+dx$ per collision from \cite{Kelner2006}. The factor $\kappa$ accounts for the changes due to different nuclei, $\kappa = \left[A_i\sigma(p,A_j, E_{p,j}) + A_j\sigma(p,A_i, E_{p,j})\right] \Bigg/ \left[2 \sigma_{\rm inel}\right]$, \cite[see][ Eq. 20]{Kafexhiu_et_al_2014}. $\sigma(p,A_j, E_{p,j})$ is the nucleus-nucleus reaction cross-section of a proton CR with a nucleus of mass number $A_j$ \citep[Eq. 18 in][]{Kafexhiu_et_al_2014}.\\

\begin{figure}
    \centering
    \includegraphics[width=0.5\textwidth]{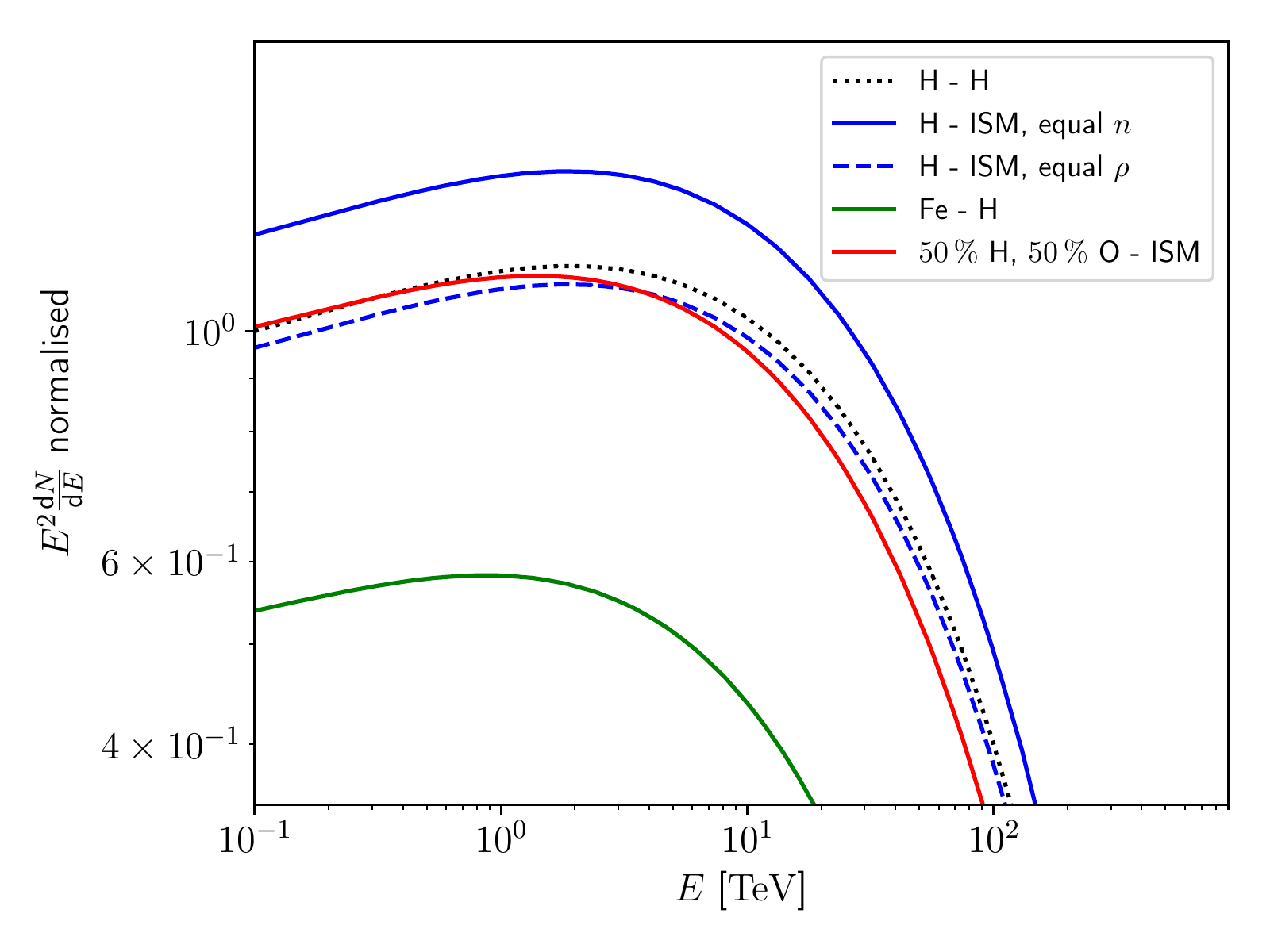}
    \caption{Gamma-ray emission for different CR and ambient medium compositions. The total energy in CRs is the same in all cases, while the cutoff in each case is rigidity dependent, set to $Z$ PeV. All curves are normalised to the value of the case of hydrogen colliding with a hydrogen ISM at \SI{100}{\giga\electronvolt} (black dotted line). The solid blue line shows the emission for the same CR spectrum but with an ISM composed of different nuclei, with the same number density as in the pure hydrogen case. In the blue dotted line, hydrogen is colliding with an ISM, but now the total ISM mass is conserved. The green line shows the spectrum from a pure iron beam colliding with hydrogen. The red line shows the case for \SI{50}{\percent} hydrogen CRs and \SI{50}{\percent} oxygen CRs colliding with the ISM (also with the same number density as in the pure hydrogen case).}
    \label{fig:figure1}
\end{figure}

To highlight the implications of composition variations, we compare in Fig. \ref{fig:figure1} the $\gamma$-ray spectral energy distributions produced by rigidity-dependent exponential cutoff power-law distributed CRs,\begin{align}
    \frac{\text d N}{\text d E} = N_{\rm p} \cdot  \left(\frac{E}{E_0}\right)^{-\alpha_{\rm p}} \exp{\left(-\frac{E}{E_{\rm cut, p}\cdot A}\right)},
    \label{eq:ECPL1}
\end{align}
for different target compositions.
All curves were normalised with respect to the idealised case of pure hydrogen (target and beam) at \SI{100}{\giga\electronvolt} and the CRs carried the same total energy above \SI{1}{\giga\electronvolt}. The cutoff for hydrogen CRs was at \SI{1}{\peta\electronvolt}, for the other species it was adjusted according to their rigidity. We used $\alpha_{\rm p}=2$ for all curves in Fig. \ref{fig:figure1}.

\begin{figure}
    \centering
    \includegraphics[width=0.5\textwidth]{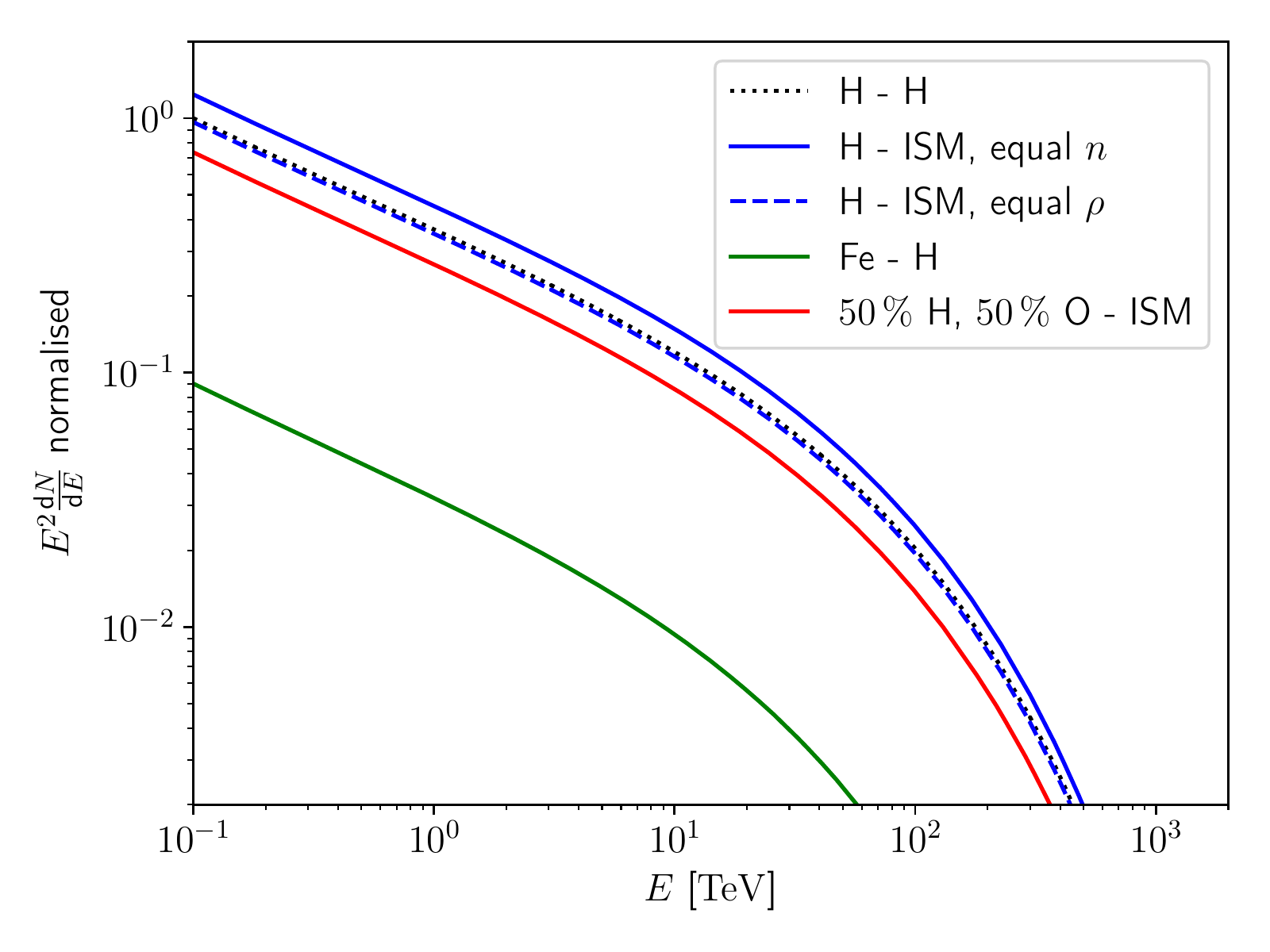}
    \caption{Same as Fig. \ref{fig:figure1}, but with a steeper power-law slope of $\alpha_{\rm p} = 2.5$.}
    \label{fig:figure2}
\end{figure}

We considered first the effect due to a different ambient medium composition. We adopted typical local Galactic ISM conditions using relative abundances of H(A=1), He(A=4), C(A=12), N(A=14), O(A=16), Ne(A=20), Mg(A=24), Si(A=28), S(A=32), and Fe(A=56) of $\num{1}:\num{9.59e-2}:\num{4.65e-4}:\num{8.3e-5}:\num{8.3e-4}:\num{1.2e-4}:\num{3.87e-5}:\num{3.69e-5}:\num{1.59e-5}:\num{3.25e-5}$ \cite[see e.g.][]{Meyer1985}. Wind or accelerator specific composition models will be left to a future study. If the total number density is conserved, relative to the reference hydrogen (i.e. pure proton-proton collisions) case, there is an approximate \SI{20}{\percent} increase in the emission  (solid blue line). This increase is constant over the whole energy range. However, if instead of the number density we fixed the total number of nucleons or the mass density in the ambient medium, there is a reduction by \SI{\sim 5}{\percent} (blue dashed line). These effects can be understood as resulting from the three-dimensional structure of the nuclei: If the number density is conserved, the nuclei are simply larger, and therefore the cross-section increases. If the number of nucleons or the mass density is conserved, the matter distribution is concentrated in the individual nuclei. The nucleons thus shield each other, reducing the cross-section slightly. 

Two other effects are observed if the hydrogen CRs are replaced by iron (green solid line). Firstly the emission in the flat part of the spectrum is noticeably reduced. A small contribution to this effect can be attributed to the normalisation; the rigidity-dependent cutoff occurs at \num{26} times that of hydrogen CRs, and since the total energy was fixed, the overall normalisation must reduce. However, the dominant effect comes from the CR composition due to nucleons shielding each other. The other crucial factor concerns the role of the spectral index of the CR distribution. In the case of a power-law particle distribution $\text dN/\text dE \propto E^{-2}$, the energy is distributed equally across the CR spectrum, which is obviously not the case for steeper or harder spectra. Because $\gamma$-ray-producing nucleons emitting at the same energy as individual hydrogen atoms reside in nuclei with energy $A$ times larger, different particle spectral shapes lead to different number densities at the required energy, which modifies the emission. This can be seen in Fig. \ref{fig:figure2}, which is equivalent in all aspects to Fig. \ref{fig:figure1} but with $\alpha_{\rm p}=2.5$. The emission from the pure iron case is one order of magnitude below the emission from pure hydrogen, a reduction factor of \num{\sim 5} compared to the $\alpha_{\rm p} = 2$ scenario.

The final effect has to do with spectral features. The cutoff in the green curve in Fig. \ref{fig:figure1} occurs at a lower energy relative to the pure hydrogen case, even though the cutoff in the CRs occurs at a higher energy due to the rigidity dependence. The reason is that although the cutoff occurs at an energy a factor of $Z$ higher, the resulting $\gamma$ rays occur at an energy $Z/A\approx 0.5$ lower, where $Z$ is the number of protons in the nucleus.

Figure \ref{fig:figure1} also shows the emission from mixed CRs, hydrogen, and oxygen, where the energy is distributed equally between both species. Although the ISM composition leads to an enhancement, the emission at \SI{100}{\giga\electronvolt} is on the same level as the pure hydrogen case. The cutoff in the $\gamma$ rays occurs at lower energies as in the hydrogen case, but not by a factor of \num{0.5}. Although in this case there is not much difference from the pure hydrogen case, the effects can be severe in other setups, especially in the case of different power-law slopes. In Fig. \ref{fig:figure2}, the differences in the red curve are already more noticeable for a slope of $\alpha_{\rm p} = 2.5$.\\

The results presented here were determined using the full GAMERA solutions. The GAMERA code is publicly available\footnote{https://github.com/libgamera/GAMERA}. 
As an additional aid for future theoretical studies, we also provide simple parametrisations for the secondary spectra produced in collisions between arbitrary power-law CR spectra with exponential cutoffs (Eq. \ref{eq:ECPL1}) and gas targets.
The resulting $\gamma$-ray and neutrino spectral energy distribution was calculated numerically and fitted with the function
\begin{align}
    f(E) = N \cdot \left(\frac{E}{E_0}\right)^{-\alpha} \exp{\left(-\left(\frac E{E_{\rm cut}}\right)^{\beta}\right)}.
    \label{eq:fitfunction}
\end{align}

Details and tabulated values for $\alpha$, $\beta$, $E_{\rm cut}$, and $N$ expressed as functions of $N_{\rm p}$, $\alpha_{\rm p}$, $E_{\rm cut,p}$, and $A$ are provided in the appendix. The parametrisation is valid until energies of $E_{\rm cut,p}\sim \SI{1}{\peta\electronvolt}$ and can therefore be used to investigate effects of the composition for high energetic individual sources and for diffuse Galactic emission from rigidity-dependent knees in the CR spectra.

\section{Galactic diffuse emission}
\label{sec:diffuse_emission}

\begin{figure*}
    \centering
    \includegraphics[width=\textwidth]{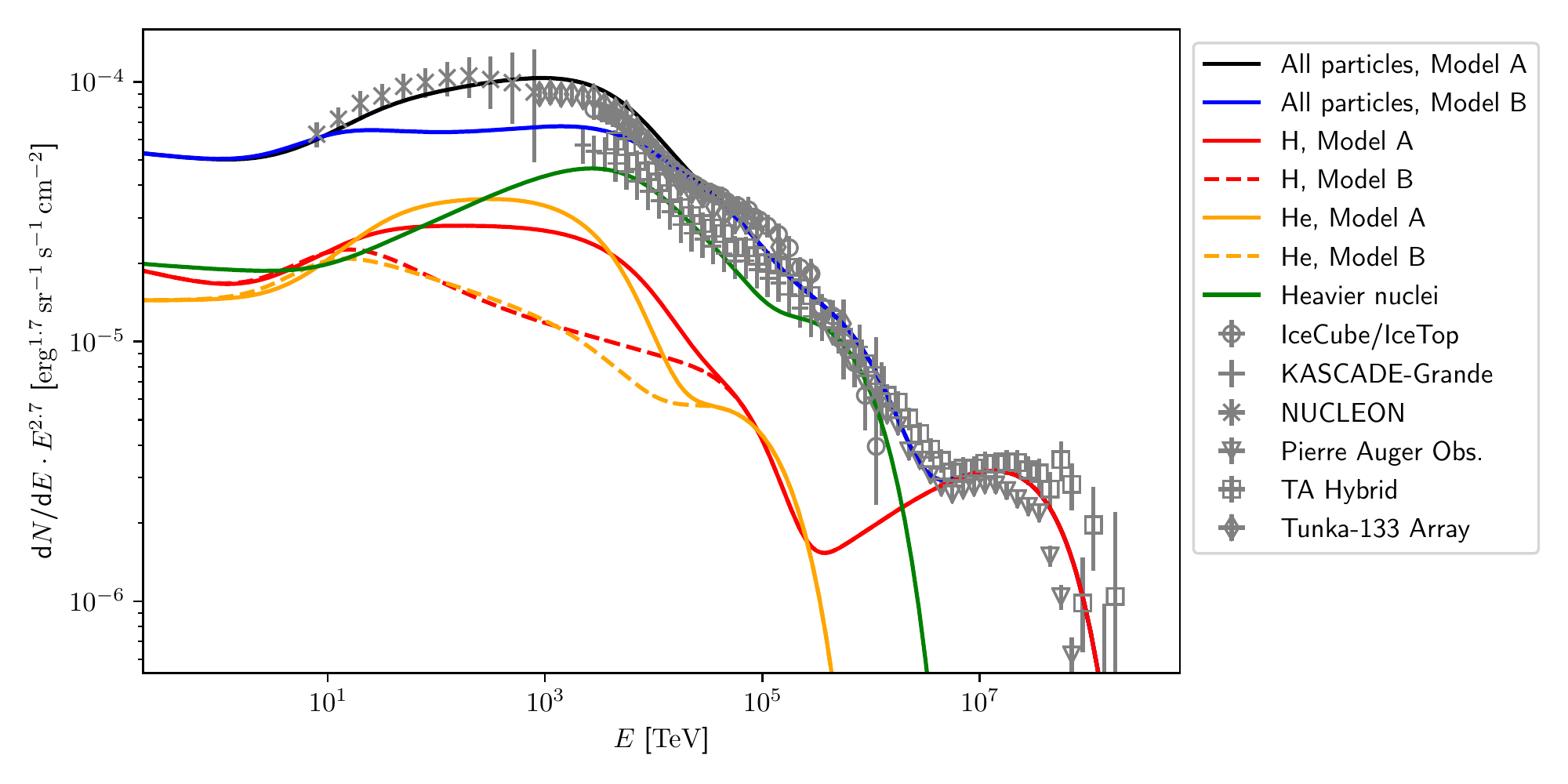}
    \caption{All-particle CR data from IceCube/IceTop \citep{IceTop_data2015}, KASCADE-Grande \citep{Kascade_data2015}, NUCLEON \citep{Nucleon_data2019}, the Pierre Auger Observatory \citep{Auger_data2015}, the Telescope Array \citep{TA_data2015}, and the Tunka-133 Array \citep{Tunka_data2014}. The black line shows the all-particle spectrum from Model A and the blue line that from Model B. We also show separately the spectra of hydrogen (red) and helium (orange), where solid lines are for Model A and dashed lines for Model B. The combined spectrum of all species heavier than H and He is depicted with the green line, which is the same for Model A as for Model B (see text for more details). The model curves for hydrogen and helium, together with the corresponding single-particle data, are displayed in Fig. \ref{fig:CR_models_H_He}. The corresponding figures for the different heavier nucleus species are shown in Appendix \ref{Appendix:CR_model}.}
    \label{fig:CR-model}
\end{figure*}

\begin{figure*}
    \centering
    \begin{subfigure}[b]{0.45\textwidth}
    \centering
    \includegraphics[width=\linewidth]{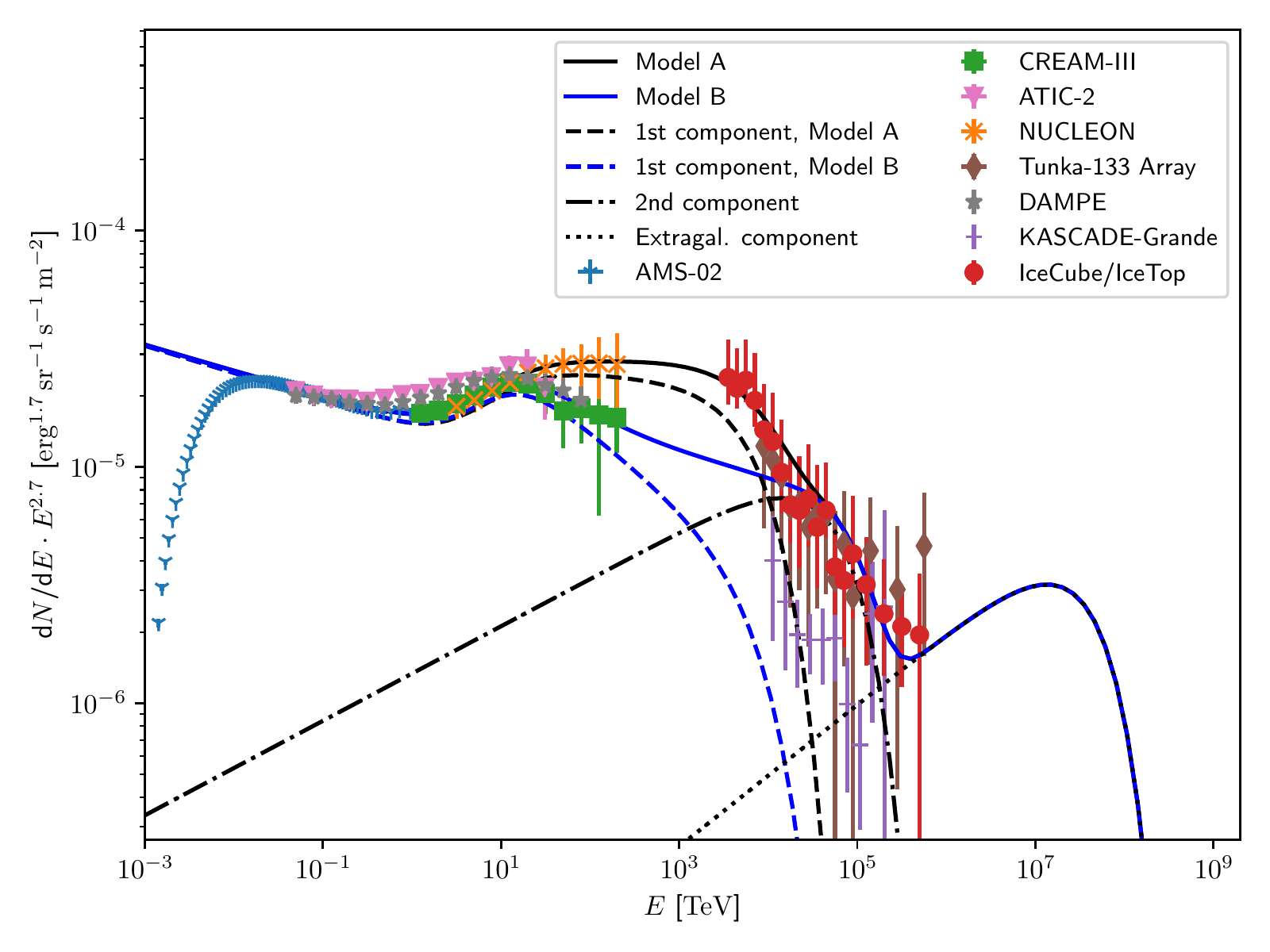}
    \caption{Hydrogen}
    \end{subfigure}
    \hspace{0.5cm}
    \begin{subfigure}[b]{0.45\textwidth}
    \centering
    \includegraphics[width=\linewidth]{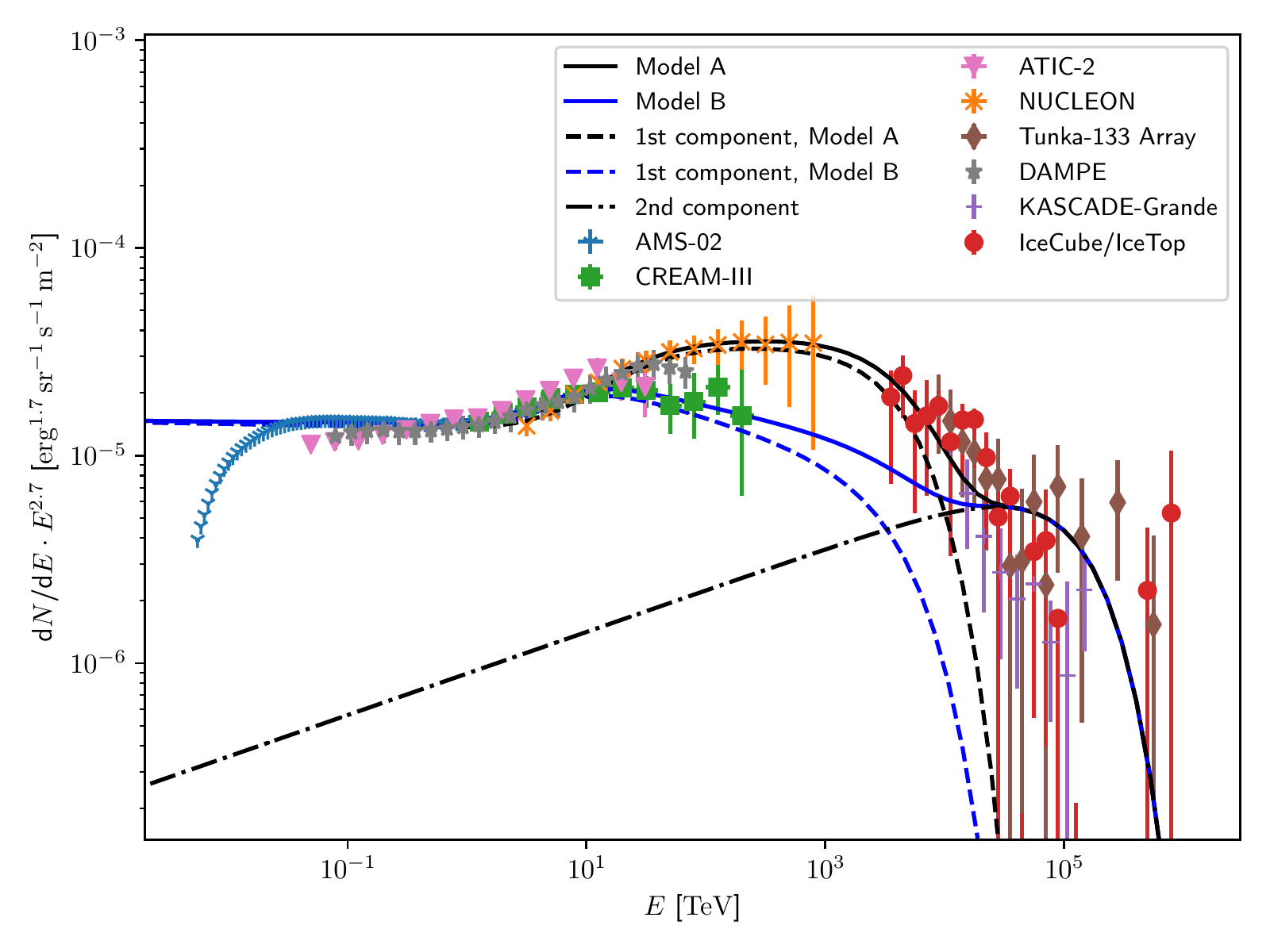}
    \caption{Helium}
    \end{subfigure}
    \caption{Cosmic-ray model components of hydrogen (left) and helium (right) together with the corresponding data from AMS-02 \citep{AMS02_2015}, CREAM-III \citep{CREAM-III_2017}, ATIC-2 \citep{ATIC_2009}, NUCLEON \citep{Nucleon_data2019}, DAMPE (\cite{DAMPE_2019_H} for H, \cite{DAMPE_2021_He} for He), Tunka-133 \citep{Tunka_data2014}, KASCADE-Grande \citep{KASCADE_data_2013}, and IceCube/IceTop \citep{IceTop_2019}. Model A is shown with the solid black line and Model B with the solid blue line. The black dashed line shows the contribution from the first Galactic component for Model A and the blue dashed line for Model B. The second components and the extragalactic component for hydrogen are the same in Models A and B and are shown with dashed-dotted lines and a dotted line in the case of the extragalactic component. Figures for the other CR species, carbon, oxygen, magnesium, silicon, and iron, can be found in Appendix \ref{Appendix:CR_model}.}
    \label{fig:CR_models_H_He}
\end{figure*}

The Galactic diffuse $\gamma$-ray emission has been measured by the Fermi-LAT Collaboration \citep{Fermi_coll_2012_diffuse_emission} between 0.1 and \SI{1000}{\giga\electronvolt}. The composition of CRs at  Earth is similarly well measured in this energy domain, and the modification of the expected diffuse $\gamma$-ray and neutrino emission relative to the proton-proton case has been calculated \citep{Mori_2009,Kachelriess_et_al_2014}. However, as shown in Sect.~\ref{sec:methodology}, the effect of a mixed composition on the diffuse emission can be more significant in the presence of cutoffs or breaks, as has also been pointed out by \citet{Kachelriess_et_al_2014}. The first measurements of the diffuse $\gamma$-ray emission between \SI{100}{\tera\electronvolt} and \SI{1}{\peta\electronvolt} were recently reported by the Tibet AS$\gamma$ Collaboration \citep{Tibet_2021}. The local CR spectrum is also well constrained in this energy domain \citep{Nucleon_data2019};  however, since spatial variations in composition and spectrum are expected \cite[see e.g.][]{Fermi_2016_diffuse}, it is also important to assess the impact of composition at these high energies.

While the improved sensitivity of the High-Altitude Water Cherenkov Gamma-Ray Observatory (HAWC) and Large High Altitude Air Shower Observatory (LHAASO) enables more precise measurements, with the data reported by the Tibet AS$\gamma$ Collaboration it is already possible to make several important conclusions. We develop in the following a simple model for the diffuse $\gamma$-ray and neutrino emission. The key elements include: a model for the CR spectrum, including shape, composition, normalisation and cutoff of the different components; gas distribution in the Galactic disk; photon field in Galactic plane to determine absorption.   
These effects have been explored previously by \citet{Lipari_Vernetto_2018} who highlight in particular the importance of absorption. They also consider variations in the spectral index across the Galaxy, and their model is used in fitting the data by the Tibet AS$\gamma$ Collaboration \citep{Tibet_2021}. Since composition effects are not discussed directly in these works, and CRs with energies above the knee are not considered, we revisit these issues here. With the new data, it is possible to better constrain the abundances of different CR species at energies of \SI{100}{\tera\electronvolt} and above.

To develop an illustrative CR model, we retrieved CR data from the Cosmic-Ray DataBase \citep{CR_database2014, CR_database2020}. We used all-particle data from IceCube/IceTop \citep{IceTop_data2015}, the Karlsruhe Shower Core and Array Detector-Grande (KASCADE-Grande) \citep{Kascade_data2015}, NUCLEON \citep{Nucleon_data2019}, the Pierre Auger Observatory \citep{Auger_data2015}, the Telescope Array (TA) \citep{TA_data2015}, and the Tunka-133 Array \citep{Tunka_data2014}. For the most abundant CR species hydrogen (H), helium (He), carbon (C), oxygen (O), magnesium (Mg), silicon (Si), and iron (Fe), we used data from the Alpha Magnetic Spectrometer-02 (AMS-02) \citep{AMS02_2015, AMS02_2017, AMS02_2020}, the Cosmic Ray Energetics and Mass-III experiment (CREAM-III) \citep{CREAM-III_2017}, CREAM-II \citep{CREAM-II_2009}, the Advanced Thin Ionization Calorimeter-2 (ATIC-2) \citep{ATIC_2009}, NUCLEON \citep{Nucleon_data2019}, the Dark Matter Particle Explorer (DAMPE) \citep{DAMPE_2019_H, DAMPE_2021_He}, Tunka-133 \citep{Tunka_data2014}, KASCADE-Grande \citep{KASCADE_data_2013}, and IceCube/IceTop \citep{IceTop_2019}.

Our total CR model consists of three different components, two Galactic components determining the shape until the ankle, and although not relevant in this study, for completeness, we included an extragalactic component dominating above $\sim$\SI{e19}{\electronvolt}. 
The first Galactic component for H and He follows an exponential cutoff power-law spectrum with two breaks:
\begin{align}
   \label{eq:double_break_ECPL}
    \frac{\text d N}{\text dE} =  N_0 \left(\frac{E}{E_0}\right)^{-\alpha_1}  &\exp\left(-\frac{E}{E_{\text{cut}}}\right)\\
    &\times ~~\left[ 1 + \left( \frac{E}{E_1}\right)^2 \right]^{\frac{-(\alpha_2 - \alpha_1)}{2}}
    \left[ 1 + \left( \frac{E}{E_2}\right)^2 \right]^{\frac{-(\alpha_3 - \alpha_2)}{2}}
  \nonumber.
\end{align}
Data from different experiments show discrepancies, especially at energies above \SI{\sim 10}{\tera\electronvolt}. To investigate the effects of different power-law slopes permitted by the data on the $\gamma$-ray and $\nu$ spectra, two different shapes for this H and He component were adopted, leading to two different models: In Model A the model follows the data from AMS-02 and NUCLEON, whereas in Model B it matches the AMS-02 and CREAM data. The values for the free parameters in Eq. \ref{eq:double_break_ECPL} can be found in Table \ref{tab:H_He_first_component}, and $E_0 = \SI{10}{\giga\electronvolt}$ in all cases.

\begin{table}[]
\caption{Parameters for the first Galactic components of H and He for Models A and B (Eq. \ref{eq:double_break_ECPL}).}
    \centering
    \begin{tabular}{|c|c c|c c|} \hline
             & \multicolumn{2}{c|}{Model A} & \multicolumn{2}{c|}{Model B} \\ \hline
    Parameter&  H            & He & H & He\\\hline
    $N_0/k$    & \num{28.2}    & \num{16.0}       & \num{28.2}  & \num{16.0}  \\ \hline 
    $E_1$ [\si{\giga\electronvolt}]  & \num{2.0e3}   & \num{4.0e3}      & \num{2.0e3}  & \num{3.0e3}  \\ \hline
    $E_2$ [\si{\giga\electronvolt}] & \num{2.5e4}   & \num{4.0e4}      & \num{1.5e4}  & \num{9.0e3}  \\ \hline
    $\alpha_1$   & \num{2.815}  & \num{2.71}  & \num{2.815}  & \num{2.71}  \\ \hline
    $\alpha_2$   & \num{2.45}   & \num{2.32}   & \num{2.4}  & \num{2.2}  \\ \hline
    $\alpha_3$   & \num{2.72}   & \num{2.66}   & \num{3.0}  & \num{2.87}  \\ \hline
    $E_{\text{cut}}$ [\si{\peta\electronvolt}] & \num{9} & \num{5}  & \num{9}  & \num{5}  \\ \hline
    \end{tabular}
    \tablefoot{$E_0$ is fixed to \SI{10}{\giga\electronvolt} in all cases. $k = 4\pi/(10^4 c) \cdot \si{\centi\meter\tothe{-2}\per\second\per\giga\electronvolt}$, where $c$ is the speed of light such that $N_0/k$ is unit-less.}
    \label{tab:H_He_first_component}
\end{table}
The first Galactic components of the elements C, O, Mg, Si, and Fe were modelled with an exponential cutoff power law with only one spectral break:\begin{align}
    \frac{\text d N}{\text dE} = N_0 \left(\frac{E}{E_0}\right)^{-\alpha_1} \left[ 1 + \left( \frac{E}{E_1}\right)^2 \right]^{\frac{-(\alpha_2 - \alpha_1)}{2}}
    \cdot \exp\left(-\frac{E}{E_{\text{cut}}}\right),
    \label{eq:broken_ECPL}
\end{align}
where $E_0$ is again fixed to \SI{10}{\giga\electronvolt}. The values of the other parameters can be found in Table \ref{tab:param_first_gal_comp}; they are the same in both models.
\begin{table}[]
    \caption{Parameters for the first Galactic components of C, O, Mg, Si, and Fe (Eq. \ref{eq:broken_ECPL}).}
    \centering
    \begin{tabular}{|c|c|c|c|c|c|}\hline
         &  $N_0/k$ &  $E_1$ [\si{\giga\electronvolt}]& $\alpha_1$ & $\alpha_2$   & $E_{\text{cut}}$ [\si{\peta\electronvolt}] \\\hline
    C & \num{3.5} & \num{7.0e3} & \num{2.7} & \num{2.45} & \num{9} \\\hline
    O & \num{6.0} & \num{7.0e3} & \num{2.7} & \num{2.5} & \num{6} \\\hline
    Mg & \num{3.0} & \num{1.0e4} & \num{2.75} & \num{2.55} & \num{42} \\\hline
    Si & \num{2.5} & \num{1.0e4} & \num{2.7} & \num{2.5} & \num{39} \\\hline
    Fe & \num{10.0} & \num{1.0e4} & \num{2.8} & \num{2.55} & \num{36} \\\hline
    \end{tabular}
    \tablefoot{$E_0$ is fixed to \SI{10}{\giga\electronvolt} in all cases. $k = 4\pi/(10^4 c) \cdot \si{\centi\meter\tothe{-2}\per\second\per\giga\electronvolt}$, where $c$ is the speed of light such that $N_0/k$ is unit-less. The parameters are the same in Model A and Model B.}
    \label{tab:param_first_gal_comp}
\end{table}

The second Galactic component follows an exponential cutoff power law for every particle species:
\begin{align}
    \frac{\text dN}{\text dE} = N_0 \left(\frac{E}{E_0}\right) \cdot \exp\left(-\frac{E}{E_{\text{cut}}}\right).
    \label{eq:ECPL}
\end{align}
It dominates between $\approx \SI{1e17}{\electronvolt}$ and $\approx \SI{3e18}{\electronvolt}$ and is responsible for the hardening of the spectrum above the second knee. In Models A and B, the parameters were kept the same; they can be found in Table \ref{tab:sec_gal_comp}. In this energy range, all-particle data as well as data for different particle groups are available. The single-particle group data in this region depend strongly on the interaction model used, and there are discrepancies between different observatories. The parameters in Table \ref{tab:sec_gal_comp} were adapted to match not only single-particle data, but also to fit the all-particle spectrum. Because we distinguished between C, O, Mg, and Si, while the data are given in terms of elemental groups, it was assumed that the data are composed of equal parts of the different group members.

\begin{table}[]
    \caption{Values for all species of the second component and the extragalactic component (Eq. \ref{eq:ECPL}).}
    \centering
    \begin{tabular}{|c|c|c|c|} \hline
        &  $N_0/k$ & $\alpha$  & $E_{\text{cut}}$ [\si{\peta\electronvolt}]  \\ \hline
    H   & \num{6.0e-6} & \num{2.5} & \num{70} \\\hline
    He  & \num{4.0e-6} & \num{2.5} & \num{140} \\\hline
    C   & \num{6.0e-7} & \num{2.45} & \num{420}\\\hline
    O   & \num{6.0e-7} & \num{2.45} & \num{560} \\\hline
    Mg  & \num{5.0e-7} & \num{2.45} & \num{840} \\\hline
    Si  & \num{3.0e-7} & \num{2.45} & \num{980} \\\hline
    Fe  & \num{8.0e-7} & \num{2.4} & \num{840} \\\hline
    H, extragalactic    & \num{1.4e-7} & \num{2.4} & \num{45e3} \\\hline
    \end{tabular}
    \tablefoot{The extragalactic component is assumed to consist solely of hydrogen for the sake of simplicity. $k = 4\pi/(10^4 c) \cdot \si{\centi\meter\tothe{-2}\per\second\per\giga\electronvolt}$, where $c$ is the speed of light such that $N_0/k$ is unit-less. The parameters are the same in Model A and Model B.}
    \label{tab:sec_gal_comp}
\end{table}

To match the hardening at the ankle above several \SI{e18}{\electronvolt}, we introduced an additional component of most likely extragalactic origin, assumed to follow an exponential cutoff power law (Eq. \ref{eq:ECPL}).
Since the composition of this component is even more speculative than the one for the second Galactic component, we made the simplifying assumption that it consists solely of hydrogen. This is motivated by evidence that the composition at the ankle is lighter; however, we emphasise that it plays no role in the results that follow. The values of the parameters of this component can be found in Table \ref{tab:sec_gal_comp} as well.\\

The CR models are illustrated in Fig. \ref{fig:CR-model}. The data points show all-particle data from IceCube/IceTop, KASCADE-Grande, NUCLEON, the Pierre Auger Observatory, the Telescope Array, and the Tunka-133 Array. The different lines show the sum of all CR species of Models A and B, the contributions of H and He for Models A and B, and all heavier species combined.
Above \si{\peta\electronvolt} energies, heavier nuclei dominate over H or He. One can see that Model B does not follow well the all-particle data from NUCLEON above \SI{\sim 10}{\tera\electronvolt}. This is because H and He in this model follow the data from CREAM, which lies below NUCLEON. Model B also undershoots the low-energy data from Tunka and IceCube/IceTop, but it is still slightly above the low-energy KASCADE-Grande data. Figure \ref{fig:CR_models_H_He} shows the model curves for H and He in detail together with the corresponding data. Not only Model A and Model B are depicted, but also the contributions from the different components. Model curves for the other individual particle species, together with the corresponding data, can be found in Appendix \ref{Appendix:CR_model}. The parameters were not only adapted to match the individual CR data but also the all-particle spectrum.

In addition to the $\gamma$ rays and neutrinos, the decay of charged pions produces $e^-$ and $e^+$, which subsequently emit $\gamma$ rays by inverse Compton (IC) scattering and bremsstrahlung. If the Galactic interstellar CR density, the density of the ISM, the large-scale radiation fields, and the effective magnetic fields do not change significantly on timescales shorter than the cooling times, the particles will have established an equilibrium between production by hadronic collisions and losses. We therefore estimated their contribution to the total $\gamma$-ray spectrum by calculating the emission from their equilibrium spectra, determined by the production rate from the CR model and losses due to synchrotron radiation, bremsstrahlung and IC scattering across the Milky Way. For these calculations we used the density model from \cite{Ferriere1998, Ferriere_et_al2007}, the magnetic field model from \cite{Jansson2012, Jansson2012_2} and the radiation model from \cite{Popescu2017} together with the cosmic microwave background (CMB). We found that the contribution was always less than a few per cent of the total emission above \SI{100}{\giga\electronvolt} and therefore negligible. Even in the worst-case scenario, in which synchrotron losses are neglected in the calculation of the equilibrium spectra, the resulting emission from secondaries is below \SI{10}{\percent}. This is consistent with the estimates of \citet{Lipari_Vernetto_2018}, and therefore, the emission from secondary particles is omitted in the following calculations.\\

At energies above \SI{\sim 100}{\tera\electronvolt}, the absorption of $\gamma$ rays by interstellar radiation fields and the CMB becomes significant. To take this into account, we calculated the absorption in the Galactic axisymmetric radiation model from \citet{Popescu2017} and the CMB. 
We assumed that the spectral shape and composition of the CRs is unchanged throughout the Galaxy but that the total CR density scales according to \citet{Lipari_Vernetto_2018} Eq. 16 with
\begin{align}
    N_{\rm CR} = N_{\odot} \cdot \frac{\sech(r/R_{\rm CR}) \sech(z/Z_{\rm CR})}{\sech(r_{\odot}/R_{\rm CR})}.
    \label{eq:CR_density}
\end{align}
Here, $N_{\odot}$ is the CR density at the Sun, $r_{\odot}$ is the radial distance of the Sun towards the Galactic centre, and $R_{\rm CR} = \SI{5.1}{kpc}$ and $Z_{\rm CR} = \SI{1}{kpc}$ are constants taken to be the same as in \citet{Lipari_Vernetto_2018}. For the density of the target material, we used the interstellar density from the model by \citet{Ferriere1998, Ferriere_et_al2007}.
The $e^+$ and $e^-$ particles produced in the $\gamma\gamma$ collisions will subsequently re-emit at lower energies. However, in the case of power-law spectra softer than $\sim -2$, this effect is not important due to the lower energy content in $\gamma$ rays in the absorption energy regime compared to the lower energies, where the additional emission occurs \citep{Murase_Beacom_2012}, and can therefore be neglected.\\

To highlight the differences induced by different particle species on the $\gamma$-ray and neutrino spectra, the composition and the spectral index were assumed to be constant throughout the Galaxy. In additional to the spatial distribution of the CRs in Eq. \ref{eq:CR_density}, we allowed for an overall normalisation constant, which was determined by fitting the $\gamma$-ray emission to the data from the Astrophysical Radiation with Ground-based Observatory at YangBaJing (ARGO-YBJ) \citep{Argo_2015_diffuse} and from the Tibet AS$\gamma$ Collaboration \citep{Tibet_2021} for the Galactic longitude range $\SI{25}{\degree} < l < \SI{100}{\degree}$ and Galactic latitudes $|b| < \SI{5}{\degree}$. For the ISM composition, the same relative abundances as in Sect. \ref{sec:methodology} were used.

\begin{figure}
    \centering
    \includegraphics[width=0.5\textwidth]{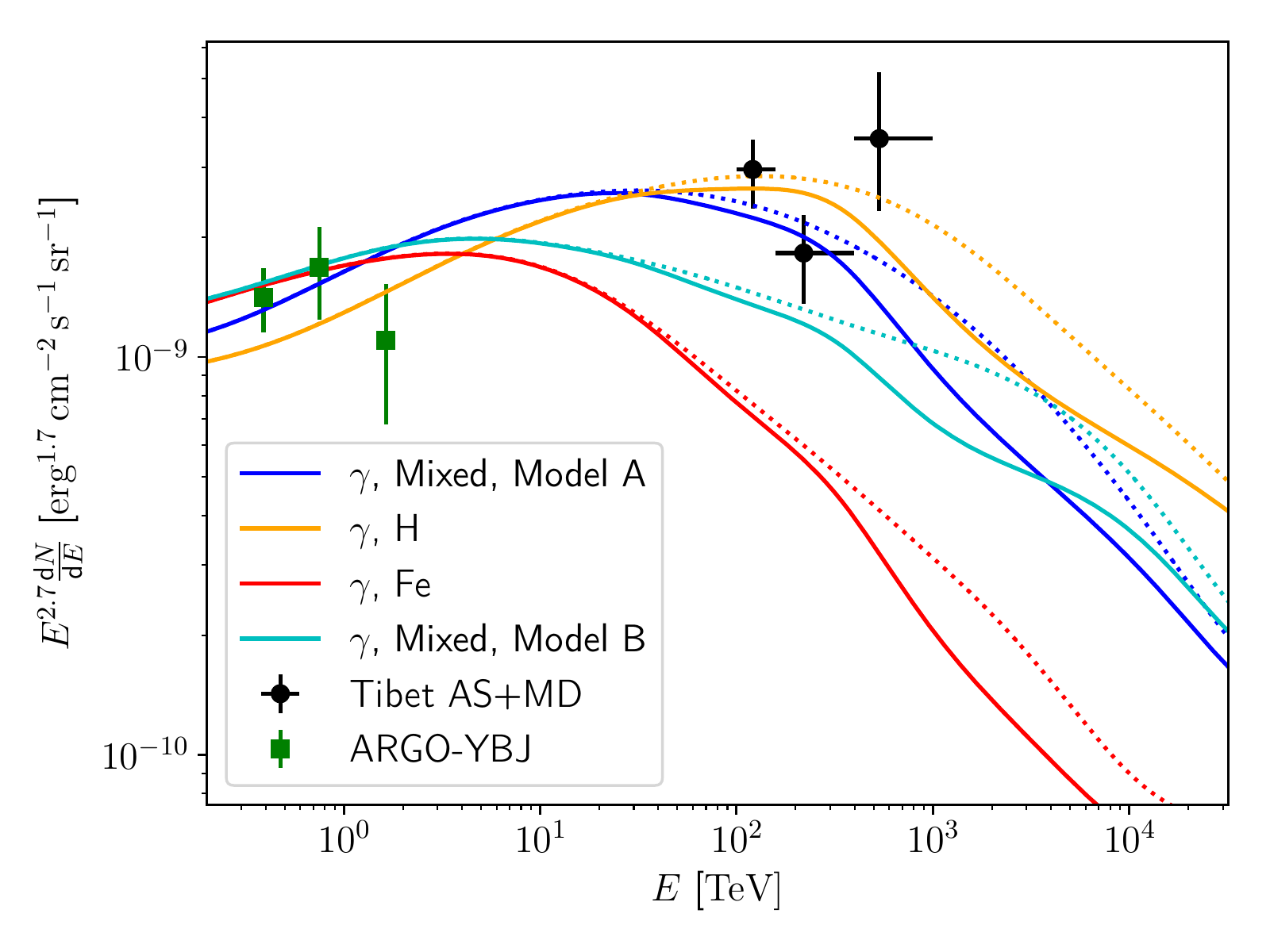}
    \caption{Gamma-ray fluxes from different CR models together with data from ARGO-YBJ \citep{Argo_2015_diffuse} and the Tibet air shower array and muon detector array (Tibet AS+MD) \citep{Tibet_2021} for Galactic longitudes $\SI{25}{\degree} \leq l \leq \SI{100}{\degree}$ and Galactic latitudes with $|b| \leq \SI{5}{\degree}$. Each colour represents a different CR model; the solid lines are with absorption, and the corresponding dotted lines without. The curves are normalised by a fit on the data by taking the absorption into account. The blue colour represents Model A, with all different CR species. The orange and red curves assume that all CR species of Model A are hydrogen or iron, respectively. The cyan curve is the emission from Model B, where the H and He CR spectra are changed with respect to Model A.
    The corresponding all-sky Galactic neutrino fluxes are shown in Fig. \ref{fig:neutrinos}.}
    \label{fig:gamma-rays}
\end{figure}

The results are shown in Fig. \ref{fig:gamma-rays}. Data from ARGO-YBJ and from the Tibet air shower and muon detector array (Tibet AS+MD) are displayed together with the emission from the mixed CR Models A and B described above and shown in Fig. \ref{fig:CR-model}. For comparison, we also plot two extreme cases where the CR particles from Model A are exclusively hydrogen or iron. The emission without absorption is shown as well.
In all cases, the normalisation was fitted separately to match best the $\gamma$-ray data.
This accounts for the similar flux levels of the pure hydrogen, the pure iron and the mixed CR models below \SI{\sim 10}{\tera\electronvolt}. The $\gamma$-ray production rate from the pure hydrogen case with the same normalisation as in the mixed model would be a factor of \num{1.9} above the mixed-composition case at \SI{1}{\tera\electronvolt}. The fitted mixed CR Model A curve is only \SI{23}{\percent} above an estimation of the flux calculated with the Galactic hydrogen density given by the model from \citet{Ferriere1998, Ferriere_et_al2007} and assuming the CR density profile in Eq. \ref{eq:CR_density} from \cite{Lipari_Vernetto_2018}.

\begin{figure}
    \centering
    \includegraphics[width=0.5\textwidth]{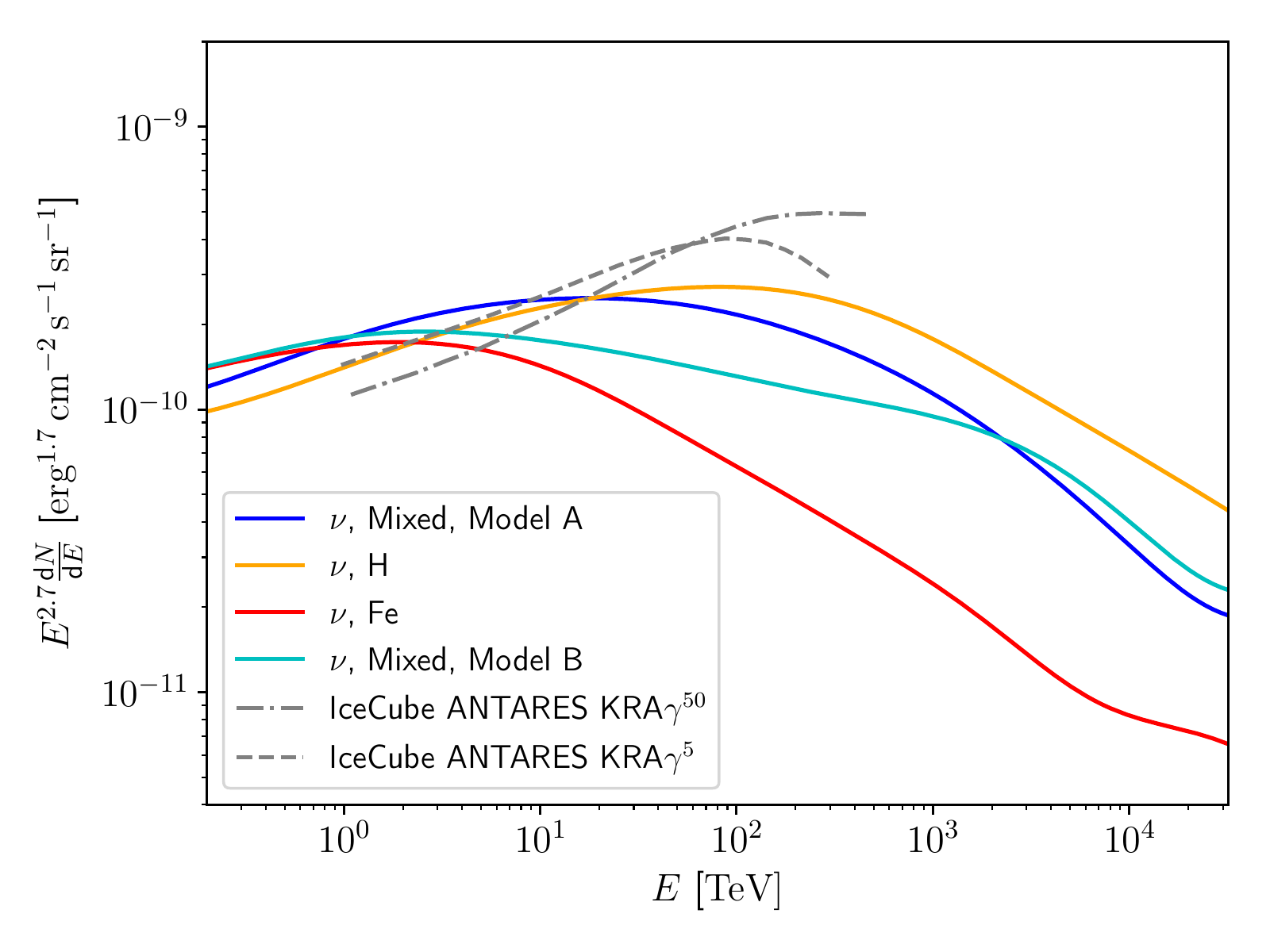}
    \caption{Predicted all-sky Galactic neutrino emission (solid lines) together with combined limits from IceCube and ANTARES \citep{IceCube_ANTARES_2018}
    (dashed and dashed-dotted line for the KRA$\gamma^{5}$ and KRA$\gamma^{50}$ model, respectively). The different colours represent neutrino emission from the same CR models for the $\gamma$-ray emission in Fig. \ref{fig:gamma-rays}. The blue line is the neutrino flux from the mixed CR Model A, and the orange and red lines show the emission assuming all CR species from Model A are H or Fe, respectively. The cyan curve shows the results for Model B.}
    \label{fig:neutrinos}
\end{figure}

Above {\SI{\sim 30}{\tera\electronvolt}}, the spectrum from the mixed CR Model {A} starts to soften with respect to the hydrogen case. This effect becomes increasingly important for higher energies. At \SI{1}{\peta\electronvolt}, the pure hydrogen case is a factor of {\num{1.5}} above the mixed {Model A}. 
For the pure Fe case, a spectral softening can be observed already below \SI{10}{\tera\electronvolt}. This is expected because any spectral feature will appear at an energy $1/A$ with respect to hydrogen, where $A=56$ for iron (see Sect. \ref{sec:methodology}). The pure iron case is of course extreme, but any different assumption about the composition will be situated between the hydrogen and the iron curves.\\

We now consider the mixed CR Model B shown in Fig. \ref{fig:gamma-rays}. In terms of the shape of the $\gamma$-ray emission, the curve lies between the mixed Model A and the pure Fe case. This demonstrates that changes in the spectral index of the underlying CR spectra can have very similar effects as a change in the composition. Disentangling these effects is nearly impossible with $\gamma$-ray data alone. Interpretations of future high-resolution measurements of the Galactic diffuse $\gamma$-ray emission should therefore take into account possible effects of a change in the composition as well as possible spectral changes.

Above energies of \SI{\sim 100}{\tera\electronvolt}, the absorption becomes noticeable. Maximum absorption occurs between \num{2} and \SI{3}{\peta\electronvolt} due to the CMB with a reduction in flux down to \SI{61}{\percent}. Because the CR density distribution is the same for all models and the composition is assumed to be equal throughout the Galaxy, the absorption is the same for all model curves and does not mask any differences due to composition. A different CR density distribution can have a moderate effect on the absorption: For the case of constant CR density but the same ISM density distribution, the flux is at most reduced down to \SI{56}{\percent}. The slightly increased absorption is caused by a larger relative fraction of the emission being emitted at further distances and therefore being subject to stronger absorption (see also Appendix \ref{appendix_absorption}).

The mixed Model A and the pure hydrogen case can match the ARGO-YBJ and the Tibet AS+MD data points best. In case of an extremely heavy composition such as in the pure Fe case, the emission drops already at energies of tens of \si{\tera\electronvolt}, making it impossible to account for the emission measured above \SI{100}{\tera\electronvolt}. The mixed Model B nearly touches the lower error of the second Tibet AS+MD data point but fails to fit the first data point. Thus, Model A is preferred over Model B, suggesting that the NUCLEON data provide a more consistent picture relative to one derived from CREAM data. A mix between Model A and the pure hydrogen case could better account for the first and the second Tibet AS+MD data point as the curves shown, but more data with more statistics are needed to derive firm conclusions. Changes in the CR spectrum could have similar effects.
None of the models can account for the highest energy data point. Two effects make this even more difficult, namely the presence of heavier nuclei and the absorption. Therefore, it seems very likely that a large contribution at this energy comes from unresolved sources. This is in line with the conclusions from the Tibet AS$\gamma$ Collaboration that four of the ten photons in that energy bin were detected within \SI{4}{\degree} of the direction of the Cygnus cocoon.
\begin{figure}
    \centering
    \begin{subfigure}{0.49\textwidth}
    \includegraphics[width=\textwidth]{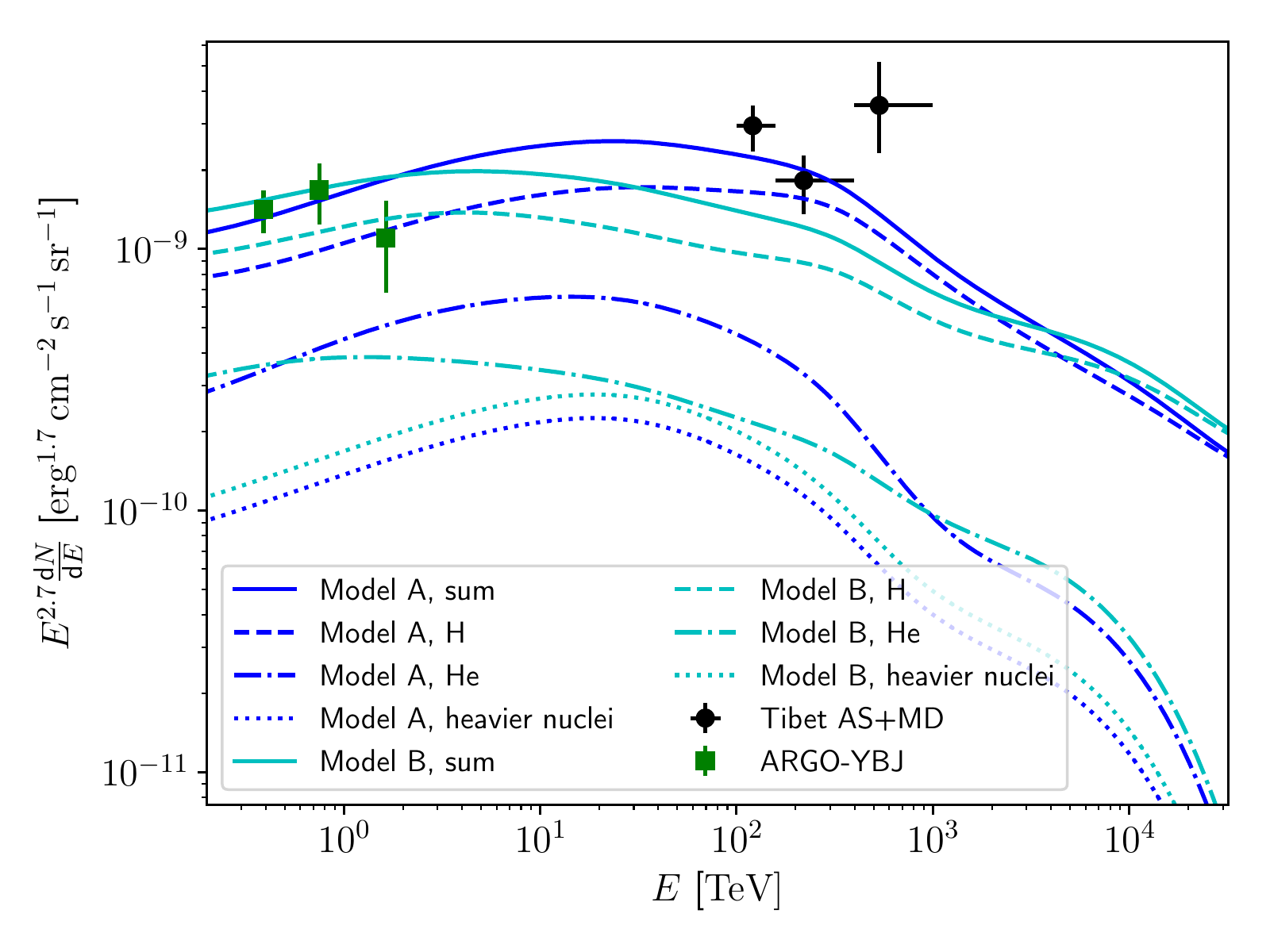}    \end{subfigure}
    \hfill
    \begin{subfigure}{0.49\textwidth}
    \includegraphics[width=\textwidth]{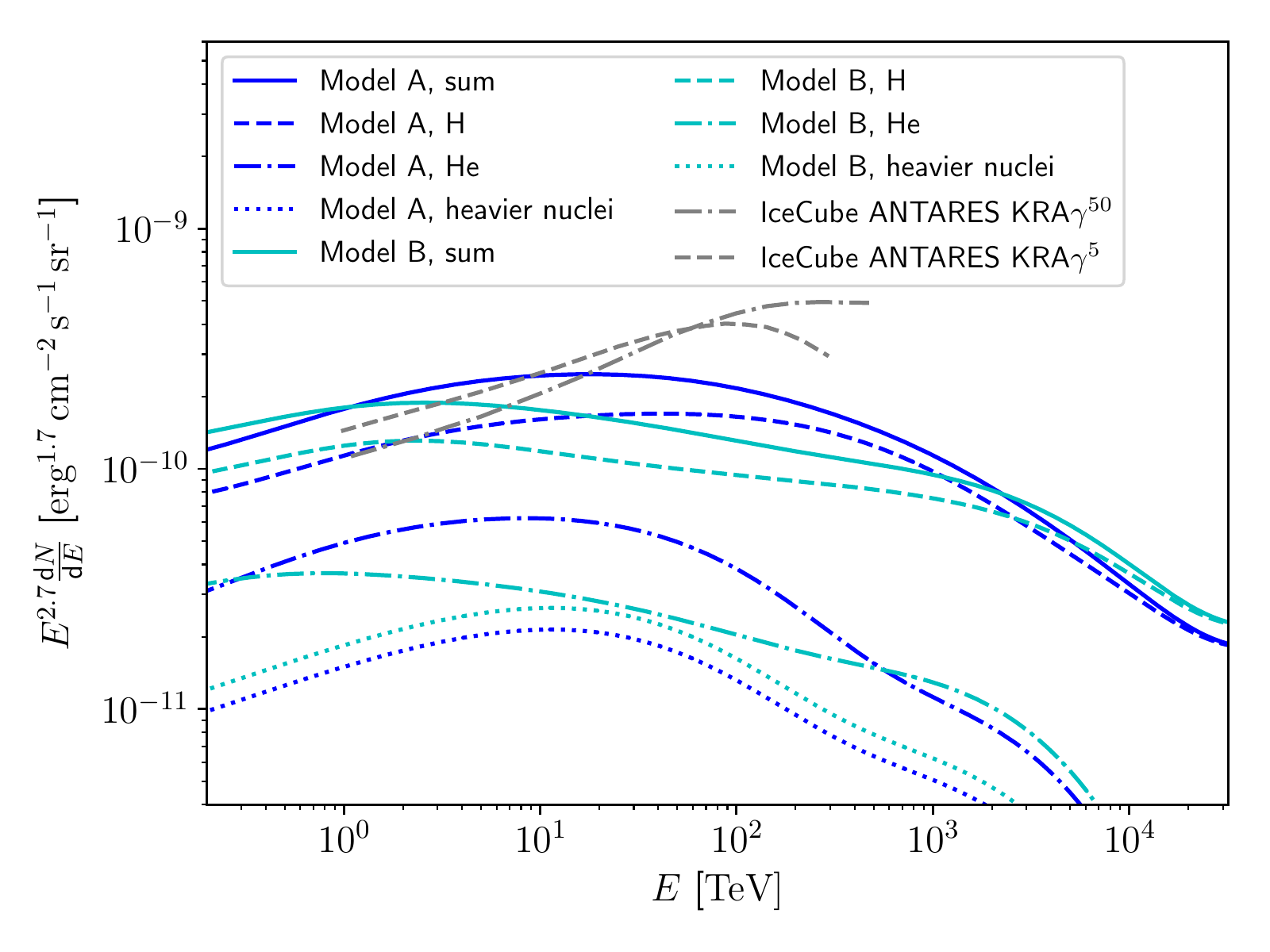}
    \end{subfigure}
    \caption{Contributions from hydrogen (dashed lines), helium (dashed-dotted lines), and heavier nuclei (dotted lines) to the total $\gamma$-ray and neutrino fluxes (solid lines). Model A is represented by blue colours and Model B by cyan. The upper panel shows the $\gamma$-ray emission together with the data from Tibet AS+MD \citep{Tibet_2021} and ARGO-YBJ \citep{Argo_2015_diffuse}, the lower panel the neutrino fluxes together with the limits from \cite{IceCube_ANTARES_2018}. The solid lines are the same as in Figs. \ref{fig:gamma-rays} and \ref{fig:neutrinos}.}
    \label{fig:contribution_different_CR_species}
\end{figure}
The upper panel of Fig. \ref{fig:contribution_different_CR_species} shows the respective contribution from H, He and combined heavier species to the $\gamma$-ray emission from both mixed CR models in Fig. \ref{fig:gamma-rays}.
Hydrogen CRs are responsible for the largest fraction of the emission ($\gtrsim\SI{64}{\percent}$ Model A, $\gtrsim\SI{67}{\percent}$ Model B), followed by helium ($\lesssim\SI{ 28}{\percent}$ Model A, $\lesssim\SI{ 24}{\percent}$ Model B). All heavier nuclei combined account for less than \SI{9}{\percent} of the emission in Model A (\SI{16}{\percent} Model B) at all energies. Changes in the fraction or spectral indices of H and He CRs will therefore have the biggest effects on the resulting $\gamma$-ray emission.

The corresponding neutrino emission was also calculated for each case and compared to combined IceCube and Astronomy with a Neutrino Telescope and Abyss Environmental Research project (ANTARES) limits from \citet{IceCube_ANTARES_2018}.
Because the limits from IceCube are for the integrated all-sky neutrino emission, the corresponding neutrino emission from the fitted models in Fig. \ref{fig:gamma-rays} was re-scaled according to the angular integrated CR and ambient medium column densities. The results are shown in Fig. \ref{fig:neutrinos}. The same trends observable in the $\gamma$ rays also appear in the neutrino fluxes. The limits presented depend strongly on the underlying models. 
At energies below \SI{\sim 10}{\tera\electronvolt}, the blue curve for Model A is slightly above the limits, but the pure hydrogen case is consistent with the KRA$\gamma^5$ limit (KRA$\gamma$ model with a \SI{5}{\peta\electronvolt} cutoff; see \citealt{Gaggero} for details). However, different assumptions in the CR and ambient medium density distribution could change this, and different underlying models in the calculation of the neutrino limits have effects as well. More statistics in the neutrino data will greatly improve the confidence of the limits. We therefore conclude that Model A and the pure hydrogen case are both broadly consistent with neutrino limits.
Future neutrino observatories might be able to probe the Galactic emission or to constrain the emission better.
The relative contributions of H, He, and heavier nuclei to the neutrino emission in both models are nearly the same as for $\gamma$ rays; the corresponding figure is shown in the lower panel of Fig. \ref{fig:contribution_different_CR_species}.
 
The results show that the effects of composition on $\gamma$-ray and neutrino emission should be taken into account in the interpretation of measurements of diffuse $\gamma$-ray emission. First, the level of emission at energies below the knee feature is affected. If the all-particle CR spectrum contains a larger amount of heavier nuclei, $\gamma$-ray and neutrino production is reduced relative to the pure hydrogen case. This can in principle be accounted for by a nuclear enhancement factor, but this factor will vary across the Galaxy. However, a change in ISM or CR density has the same effect and would be difficult to disentangle. 
The second effect related to composition concerns the shape of the break associated with the knee: if heavier nuclei become more abundant, this break will be shifted to lower energy if the energy at which the knee occurs remains the same. In theory, this allows measurement of the composition independently across the Galaxy if sufficiently accurate data are available and the CR spectrum is unchanged. If the CR knee is not constant across the  Galaxy, for example close to extreme CR sources, or the spectral shape of the CRs changes throughout the Galaxy, it would have similar effects.
Future measurements should be interpreted with these different effects in mind, and the composition should not be neglected.

\section{Conclusions}
\label{sec:conclusion}

The mixed composition of CRs and target gas is frequently neglected in $\gamma$-ray studies. A notable exception is \cite{Ahlers_et_al_2016}, in which a strong dependence on composition was demonstrated for estimates of the diffuse neutrino flux. The sensitivity of the neutrino flux to the CR composition was also explored by \cite{Joshi_Winter_Gupta_2014}, revealing a similar behaviour. We have shown that for realistic composition and the expected rigidity-dependent cutoffs, there is a significant softening expected in the spectra of hadronic sources and the diffuse Galactic emission from strong interactions. The mixed composition enhances the CR knee as a feature of diffuse $\gamma$-ray emission, with a steeper decline expected above \SI{\sim 100}{\tera\electronvolt} that makes the feature easier to detect. This implicitly assumes that the composition and spectrum of CRs measured at Earth are characteristic of the Galaxy as a whole. Such studies are timely given the reported differences between the hydrogen spectra above \SI{\sim 1}{\peta\electronvolt} calculated by the KASCADE and IceCube/IceTop experiments \cite[see][for discussion]{LV20}.

The recent Tibet AS+MD measurement of Galactic diffuse emission at hundreds of \si{\tera\electronvolt} just starts to test this paradigm, but near-future measurements with LHAASO and later SWGO will tightly constrain not just the spectrum of CRs throughout the Galaxy, but also the composition around the knee and beyond. Similarly, current neutrino limits are very close to the expected fluxes \citep{IceCube_2017_galactic_diffuse_limits} given the paradigm of a universal spectrum and composition. Next-generation instruments will certainly detect this emission and strongly complement the $\gamma$-ray measurements.

We note that $\gamma$-ray instruments with greater sensitivity and resolution (in comparison to Tibet AS+MD) are required to resolve out the majority of the source population, in particular given the new paradigm that UHE $\gamma$-ray emission seems to be rather common in, for example, pulsar wind nebulae~\citep{HAWCUHEpulsars, Breuhaus21}. 

Given the prevalence in the community of simplified treatments for mixed composition (when considered at all) that do not properly capture breaks and cutoffs in the parent population, we expect the parameterisations provided here to be of rather general use in improving predictions for VHE and UHE $\gamma$-ray and neutrino emission.

\section{Acknowledgements}
The authors thank Prof. F. Aharonian and Dr. G. Giacinti for valuable discussions. MB thanks Dr. J. Hahn and Dr. Carlo Romoli for advice on GAMERA \citep{gamera}. We also used the python packages SciPy \citep{SciPy}, NumPy \citep{NumPy} and Matplotlib \citep{Matplotlib}.

We note that in finalising this work, a number of pre-prints related to the results of \citet{Tibet_2021}  have been made public
\citep{Dzhatdoev21, LW21,Qiao_et_al_2021, Fang_Murase_2021, KNS21, Serpico21, Maity_Laha_et_al_2021, Zhang_et_al_2021}. The work presented here was carried out independently. Our results are complementary to several of these studies.

\bibliographystyle{aa}
\bibliography{Nuclei}

\onecolumn

\begin{appendix}

\section{Comparison of different CR species}
\label{sec:appendix1}
\begin{figure}[!h]
    \centering
    \includegraphics{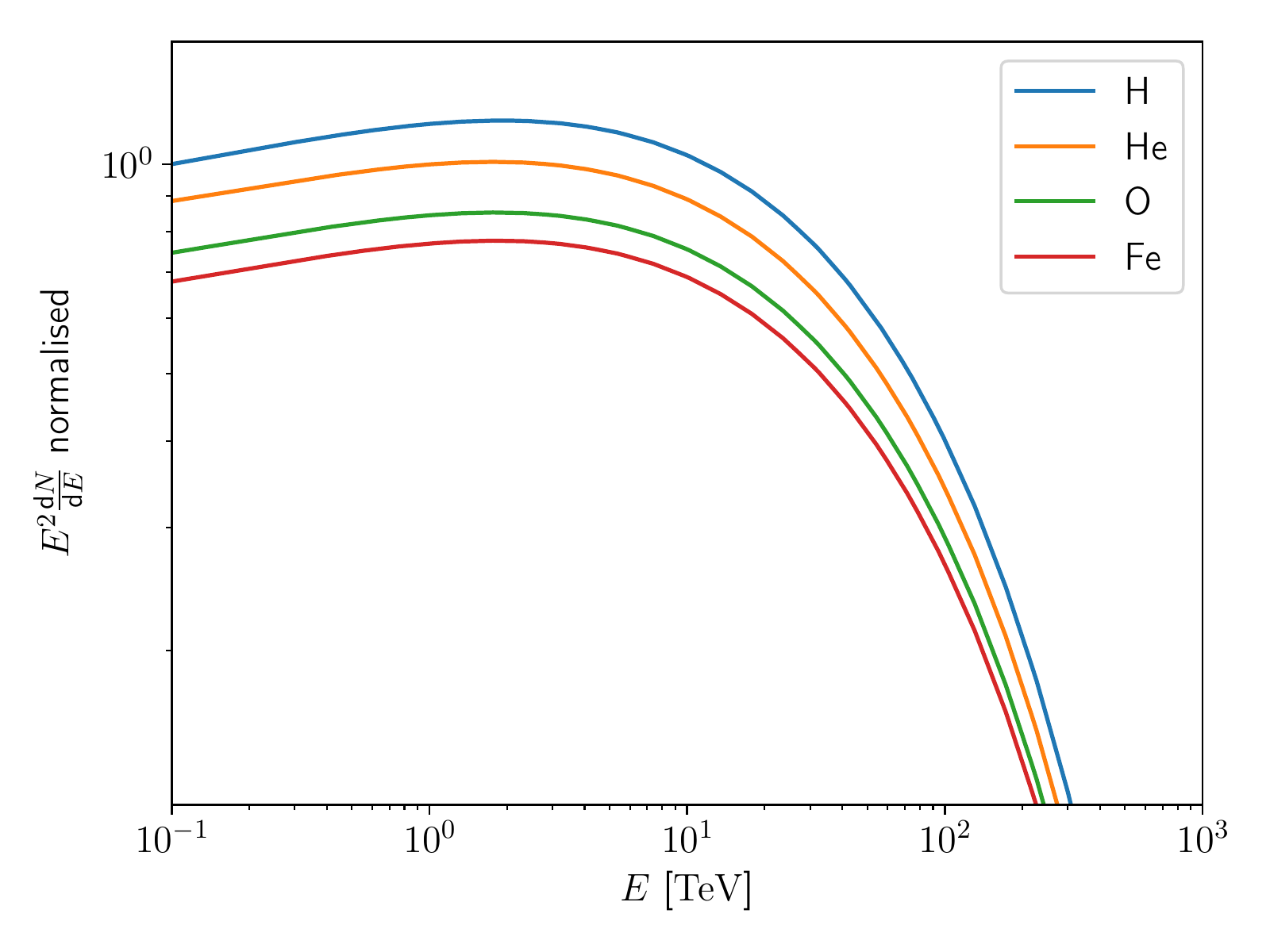}
    \caption{Gamma-ray spectra resulting from interactions of exponentially cut-off power-law distributions. To highlight the effect of shielding, here the same number of nucleons in a given energy range are present for each species, only distributed into the different particle nuclei: hydrogen, helium, oxygen, and iron.}
    \label{fig:shielding}
\end{figure}

In Sect. \ref{sec:methodology} we discussed the impact of different species on the $\gamma$-ray and neutrino spectra for sources with a rigidity-dependent cutoff. Here, we elaborate on one of the effects mentioned: the reduction due to shielding within nuclei. The effect is illustrated in Fig. \ref{fig:shielding}. We let the same nucleons produce the emission, but they are in the first case freely flying as proton CRs, and in the other cases packed into helium, oxygen, and iron atoms. The cutoff in the $\gamma$-ray spectra manifests at the same energy, but the overall emission for iron is reduced to less than \SI{70}{\percent} relative to hydrogen atoms. This effect is almost independent of the power-law slope of the CRs.

\FloatBarrier

\section{Parametrisation}
\label{sec:parametrisation}

As discussed in the main part of the text, we provide parametrised fits for the emitted spectra of $\gamma$ rays and neutrinos for exponential cutoff power-law distributions of the form
\begin{align}
    \frac{\text d N}{\text d E} = N_{\rm p} \cdot  \left(\frac{E}{E_0}\right)^{-\alpha_{\rm p}} \exp{\left(-\frac{E}{E_{\rm cut, p}\cdot A}\right)},
    \label{eq:SM1}
\end{align}
where $A$ is the mass number of the beam.

For the target material, we considered the same mixed composition described in Sect. \ref{sec:methodology}. The $\gamma$-ray and neutrino spectral energy distributions were calculated numerically and fitted with the function
\begin{align}
    f(E) = N \cdot \left(\frac{E}{E_0}\right)^{-\alpha} \exp{\left(-\left(\frac E{E_{\rm cut}}\right)^{\beta}\right)},
\end{align}
with reference energy $E_0 = \SI{1}{\tera\electronvolt}$. We used the following representations for $N$, $\alpha$, $E_{\rm cut}$, and $\beta$ as functions of $N_{\rm p}$, $\alpha_{\rm p}$, $E_{\rm cut,p}$, and $A$:

\begin{align}
\beta &= \sum_{i=0}^1 a_i \alpha_{\rm p}^i \ln (b_i E_{\rm cut,p}) + \alpha_{\rm p}^2(c_3 E_{\rm cut,p}^{d_3}+ e_3),  \nonumber \\
\alpha &= \alpha_{\rm p} - \delta - 2 \qquad \text{with} \nonumber\\
\delta &= \sum_{i=0}^2 \alpha_{\rm p}^i (c_i E_{\rm cut,p}^{d_i} + e_i), \nonumber \\
E_{\rm cut} &= 1000 \cdot E_{\rm cut,p} \cdot \sum_{i=0}^{2} \alpha_{\rm p}^i \left(
        \sum_{j=0}^{3} k_{ij} \ln^j(E_{\rm cut,p}) \right), \label{eq:fit_params}\\
    N &= \frac{N_{\rm p}\cdot \SI{1}{erg^{2}}}{10^{17}} \cdot 10^{\kappa}, \qquad \text{with} \nonumber\\
    \kappa &= p_0 \cdot \alpha_{\rm p}^2+ p_1\cdot A^{p_2} + p_3\cdot A^{p_4} \cdot \alpha_{\rm p} \qquad \text{and} \nonumber\\
    p_i &= q_{i0} \cdot E_{\rm cut,p}^{q_{i1}} + q_{i2}. \nonumber
\end{align}

The numerically determined values for these variables, for both $\gamma$ rays and neutrinos, are tabulated below. While we were free to choose $E_{\rm cut,p}$, its unit was set to \si{\peta\electronvolt}, as the value most relevant to the present study. In determining the final $\gamma$-ray and neutrino spectra, Eq. \ref{eq:fitfunction} should be multiplied with the total number density of the ambient medium.

These approximate parametrisations provide a good description for $\gamma$-ray energies between \SI{0.1}{\tera\electronvolt} and $E_{\rm cut, p}$ with the free parameters $1.8 \leq \alpha_{\rm p} \leq 3.0$, $1 \leq A \leq 56$ and $\SI{10}{\tera\electronvolt} \leq E_{\rm cut, p} \leq \SI{1}{\peta\electronvolt}$. It works best, with an accuracy much better than \SI{30}{\percent} in most cases, for $\alpha \in [2.0,3.0]$, $\gamma$-ray energies below $0.5\times E_{\rm cut, p}$, and neutrino energies below $0.2\times E_{\rm cut, p}$.\\

At energies well above the cutoff energy in the respective $\gamma$-ray and neutrino spectra, the accuracy of the parametrisation degrades rapidly. But because the emission above these energies is strongly suppressed compared to the level before the cutoff, this is not a major issue for applications such as predicting the emission from sources with different compositions or fitting spectra to data. Since the cutoff in the produced neutrinos appears at lower energies than for $\gamma$ rays produced by the same CRs, the validity range for the parametrisation of the production of neutrinos is shifted to lower energies.\\

\begin{figure}[!h]
    \centering
    \includegraphics{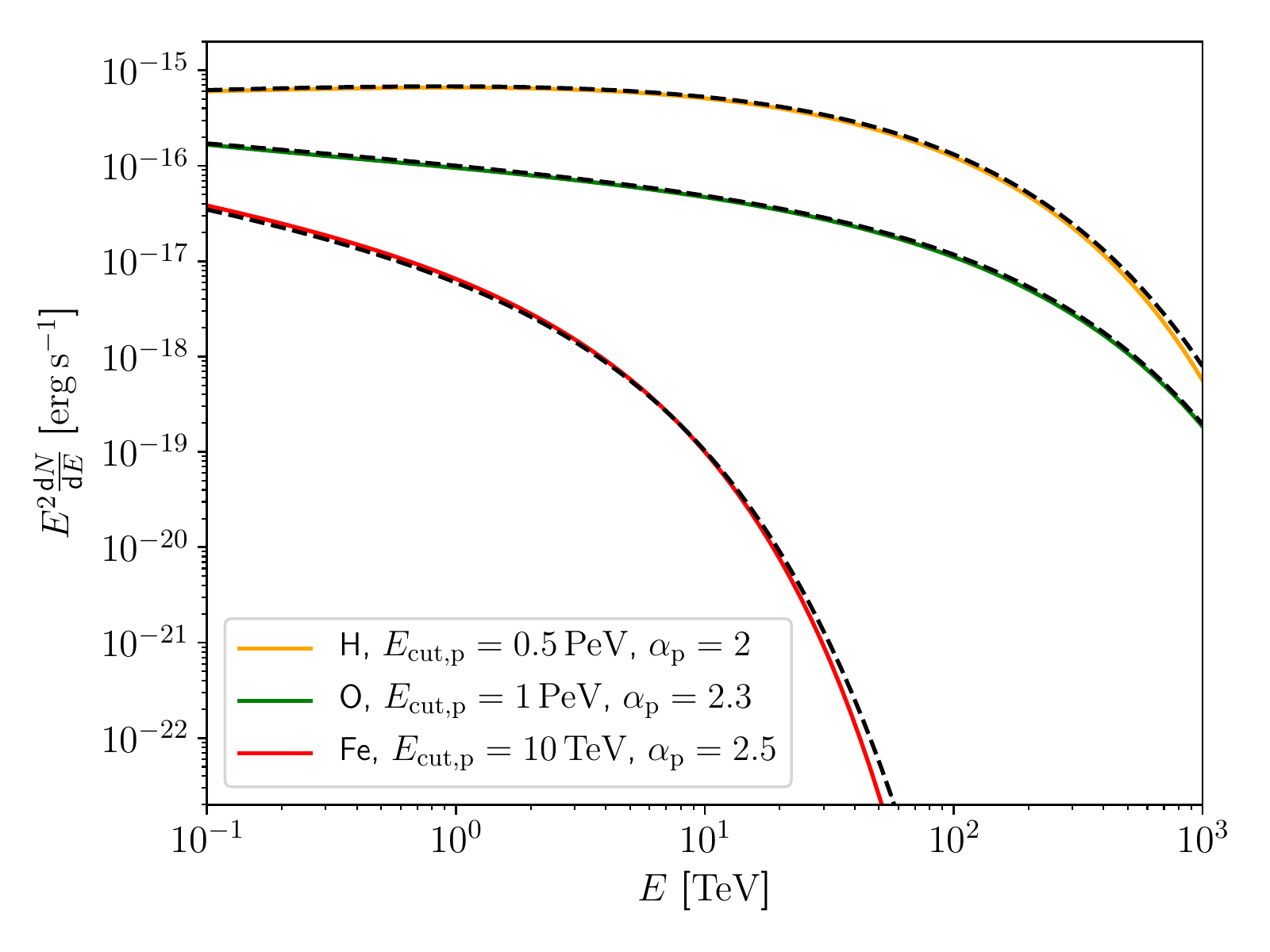}
    \caption{Gamma-ray spectra for the CR species hydrogen, oxygen, and iron for different values of $E_{\rm cut,p}$ and $\alpha_{\rm p}$ in Eq. \ref{eq:SM1}. The black dashed lines show the corresponding parametrisations. In all cases, $N_{\rm p}=1$ and $n=1$, where $n$ is the ambient density.}
    \label{fig:parametrisation}
\end{figure}

In Fig. \ref{fig:parametrisation} the parametrisation for the production of $\gamma$ rays is compared to the full GAMERA calculations for the CR species hydrogen, oxygen and iron for arbitrary values of $E_{\rm cut,p}$ and $\alpha_{\rm p}$.

\begin{table}[h!]
    \caption{Parameters $a_i$ and $b_i$ used for the calculation of $\beta$ for $\gamma$ rays and neutrinos in Eq. \ref{eq:fit_params}.}
    \centering
    \begin{tabular}{c|c|c|c|c}
        & \multicolumn{2}{c|}{$\gamma$ rays} &  \multicolumn{2}{c}{Neutrinos}\\ \hline
         &  $a_i$   &   $b_i$    &  $a_i$   &   $b_i$   \\\hline 
    $i=0$& \num{-9.23e-02}& \num{1.93e+01} & \num{-9.83e-02} & \num{4.48} \\\hline
    $i=1$& \num{7.10e-02}& \num{1.16e+03} & \num{6.82e-02} & \num{2.65e+02} \\\hline    
    \end{tabular}
    \label{tab:a_b}
\end{table}

\begin{table}[h!]
    \caption{Parameters $c_i$, $d_i$, and $e_i$ for the calculation of the parameter $\delta$ (and $\beta$) for $\gamma$ rays in Eq. \ref{eq:fit_params}.}
    \centering
    \begin{tabular}{|c|c|c|c|}
    \multicolumn{4}{c}{$\gamma$ rays} \\ \hline
         &  $c_i$   & $d_i$ & $e_i$\\ \hline
    $i=0$& \num{-4.91e-02}   & \num{-4.81e-01}   & \num{3.00e-01}   \\\hline
    $i=1$& \num{4.78e-02}   & \num{-4.19e-01}   & \num{-1.51e-01}   \\\hline
    $i=2$& \num{-1.25e-02}   & \num{-3.63e-01}   & \num{2.78e-02}   \\\hline
    $i=3$& \num{-1.21e-01}   & \num{1.52e-01}   & \num{6.47e-02}   \\\hline
    \end{tabular}
    \label{tab:c_d_e_gamma-rays}
\end{table}

\begin{table}[h!]
    \caption{Parameters $c_i$, $d_i$, and $e_i$ for the calculation of the parameter $\delta$ (and $\beta$) for neutrinos in Eq. \ref{eq:fit_params}.}
    \centering
    \begin{tabular}{|c|c|c|c|}
    \multicolumn{4}{c}{Neutrinos} \\ \hline
         &  $c_i$   & $d_i$ & $e_i$\\ \hline
    $i=0$& \num{-2.60e-01}   & \num{-2.86e-01}   & \num{6.01e-01}   \\\hline
    $i=1$& \num{3.61e-01}   & \num{-1.96e-01}   & \num{-4.87e-01}   \\\hline
    $i=2$& \num{-1.03e-01}   & \num{-1.45e-01}   & \num{1.18e-01}   \\\hline
    $i=3$& \num{-1.25e-01}   & \num{1.22e-01}   & \num{8.59e-02}   \\\hline
    \end{tabular}
    \label{tab:c_d_e_neutrinos}
\end{table}

\begin{table}[h!]
    \caption{Table for the parameters $k_{ij}$ for the calculation of $E_{\rm cut}$ in Eq. \ref{eq:fit_params} for $\gamma$ rays.}
    \centering
\begin{tabular}{|c|c|c|c|c|}
    \multicolumn{5}{c}{$\gamma$ rays} \\ \hline
    $k_{ij}$&   $j=0$   & $j=1$     & $j=2$     & $j=3$ \\ \hline    $i=0$   &\num{-1.77e-02}&\num{-5.32e-03}& \num{5.46e-03} &  \num{8.50e-04} \\ \hline
    $i=1$   & \num{-3.06e-02}&\num{7.91e-03}& \num{-5.54e-03}  &\num{-1.07e-03} \\ \hline
    $i=2$   & \num{2.89e-02}  & \num{-3.53e-03}  &  \num{1.17e-03} &\num{2.81e-04}\\ \hline
\end{tabular}
\label{tab:K_ij_gammas}
\end{table}

\begin{table}[h!]
    \caption{Table for the parameters $k_{ij}$ for the calculation of $E_{\rm cut}$ in Eq. \ref{eq:fit_params} for neutrinos.}
\centering
\begin{tabular}{|c|c|c|c|c|}
    \multicolumn{5}{c}{Neutrinos} \\ \hline
    $k_{ij}$&   $j=0$   & $j=1$     & $j=2$     & $j=3$ \\ \hline    $i=0$   &\num{3.10e-02}&\num{-3.98e-03}& \num{5.02e-04} &  \num{1.50e-04} \\ \hline
    $i=1$   & \num{-5.44e-02}&\num{3.48e-03}& \num{-5.34e-04}  &\num{-2.45e-04} \\ \hline
    $i=2$   & \num{2.30e-02}  & \num{-9.52e-04}  &  \num{7.02e-05} &\num{5.68e-05}\\ \hline
\end{tabular}
\label{tab:k_ij_neutrinos}
\end{table}

\begin{table}[h!]
    \caption{Parameters $q_{ij}$ for the calculation of the parameters $p_i$ for $\gamma$ rays in Eq. \ref{eq:fit_params}.}
    \centering
    \begin{tabular}{|c|c|c|c|}
    \multicolumn{4}{c}{$\gamma$ rays} \\ \hline
    $q_{ij}$&  $j=0$   & $j=1$     &$j=2$   \\ \hline
    $i=0$& \num{-2.69e-02}&\num{-4.54e-01}&\num{4.99e-01} \\ \hline
    $i=1$& \num{-1.02e-01}   & \num{-5.51e-01}    &   \num{7.16} \\ \hline
    $i=2$& \num{8.74e-04}   & \num{-6.24e-01}    &   \num{9.20e-02} \\ \hline
    $i=3$& \num{1.09e-01}   & \num{-4.91e-01}    &   \num{-3.62} \\ \hline
    $i=4$& \num{1.67e-03}   & \num{-6.13e-01}    &   \num{9.61e-02} \\ \hline
    \end{tabular}
    \label{tab:q_gamma-rays}
\end{table}

\begin{table}[h!]
     \caption{Parameters $q_{ij}$ for the calculation of the parameters $p_i$ for neutrinos in Eq. \ref{eq:fit_params}.}
    \centering
    \begin{tabular}{|c|c|c|c|}
    \multicolumn{4}{c}{Neutrinos} \\ \hline
    $q_{ij}$&  $j=0$   & $j=1$     &$j=2$   \\ \hline
    $i=0$& \num{-1.14e-01}& \num{-2.97e-01}& \num{6.11e-01} \\ \hline
    $i=1$& \num{-3.68e-01}   & \num{-4.12e-01}    &   \num{8.21} \\ \hline
    $i=2$& \num{2.13e-03}   & \num{-5.46e-01}    &   \num{8.31e-02} \\ \hline
    $i=3$& \num{4.38e-01}   & \num{-3.42e-01}    &   \num{-4.34} \\ \hline
    $i=4$& \num{3.99e-03}   & \num{-5.61e-01}    &   \num{8.59e-02} \\ \hline
    \end{tabular}
    \label{tab:q_neutrinos}
\end{table}

\FloatBarrier

\section{CR model components for heavier nuclei}
\label{Appendix:CR_model}
\begin{figure}[!h]
    \centering
    \begin{subfigure}{0.49\textwidth}
    \centering
    \includegraphics[width=\linewidth]{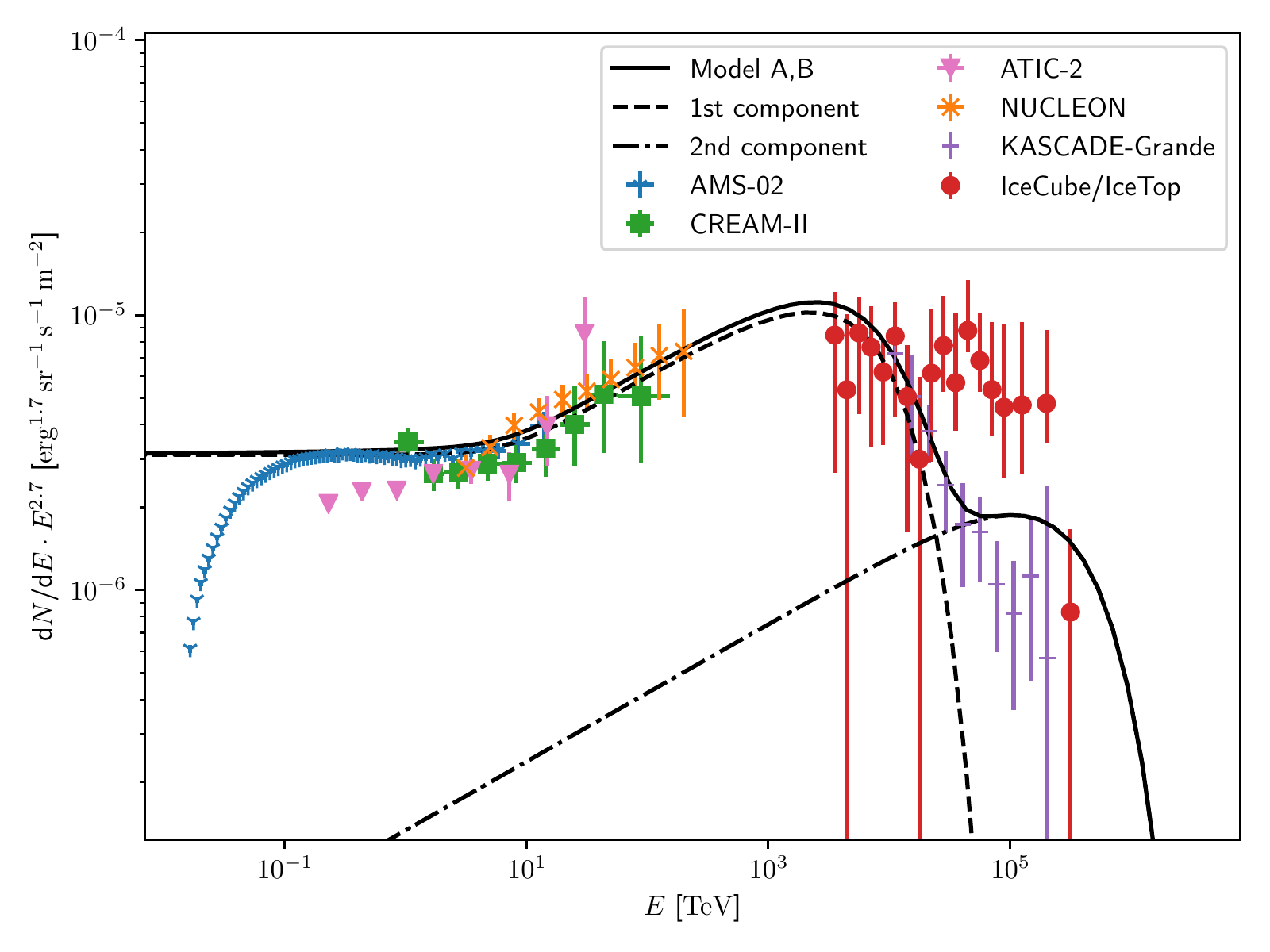}
    \caption{Carbon}
    \label{fig:CR_model_C}
    \end{subfigure}
    \begin{subfigure}{0.49\textwidth}
    \centering
    \includegraphics[width=\linewidth]{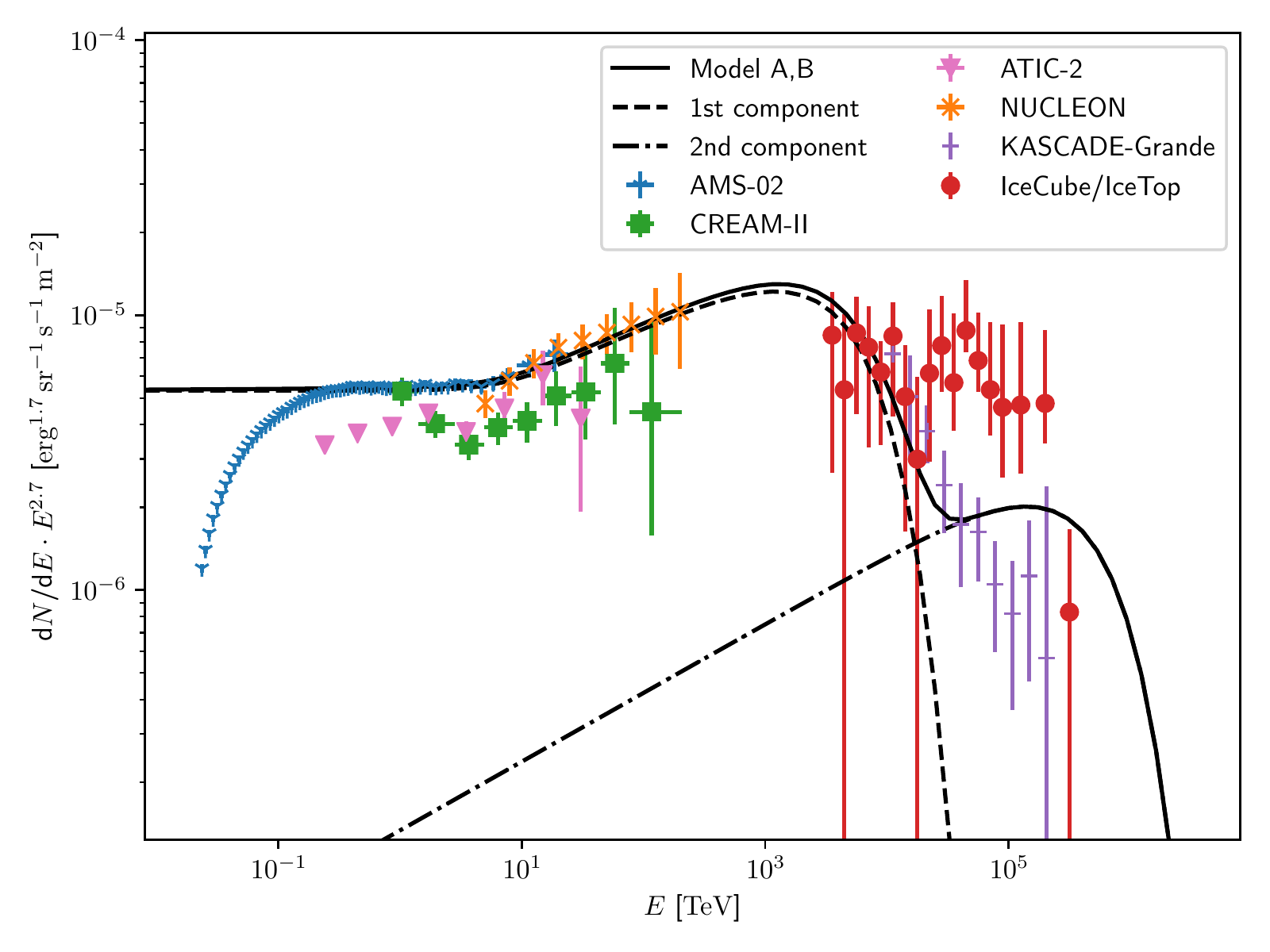}
    \caption{Oxygen}
    \label{fig:CR_model_O}
    \end{subfigure}
    \begin{subfigure}{0.49\textwidth}
    \centering
    \includegraphics[width=\linewidth]{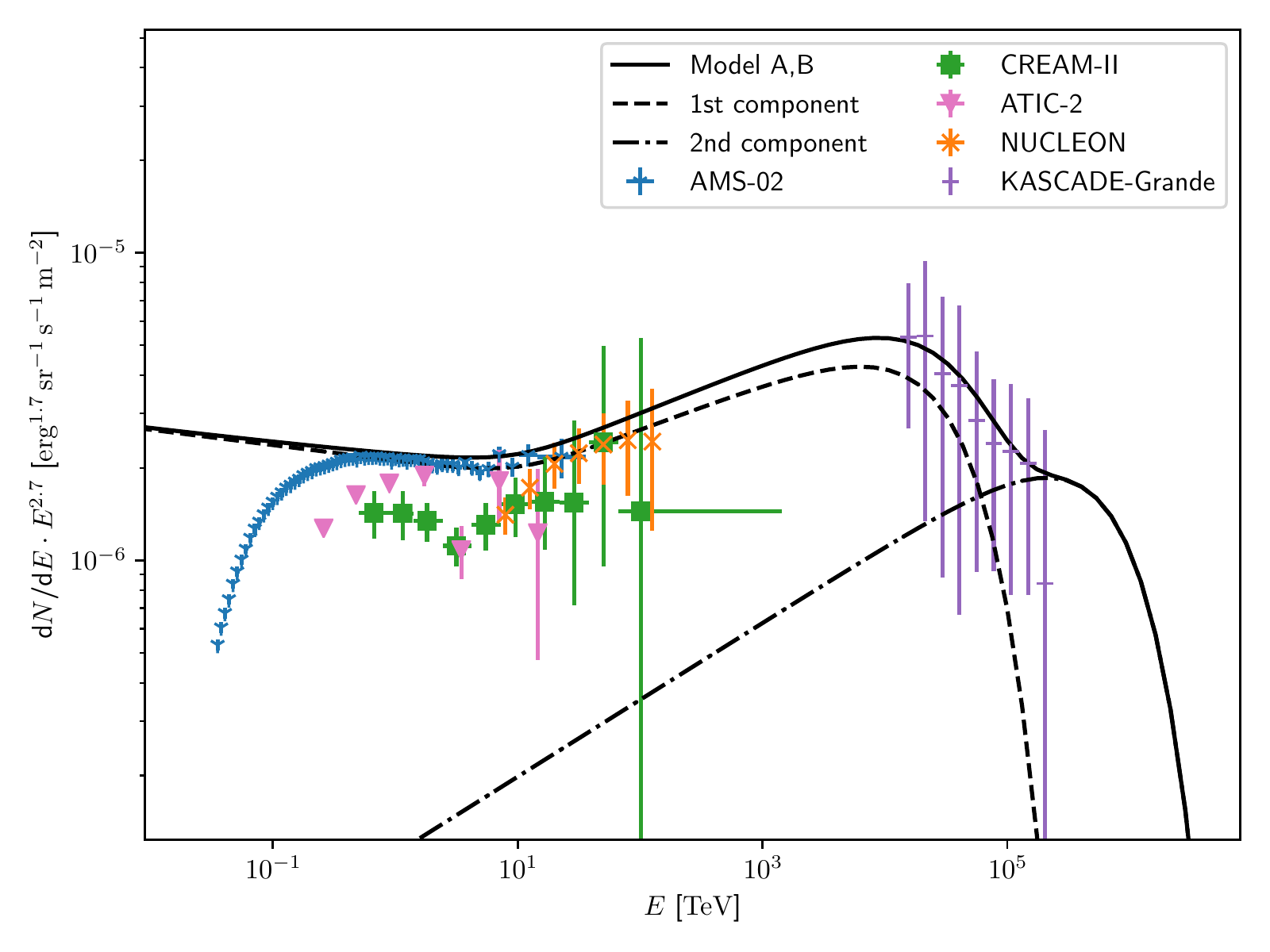}
    \caption{Magnesium}
    \label{fig:CR_model_Mg}
    \end{subfigure}
    \begin{subfigure}{0.49\textwidth}
    \centering
    \includegraphics[width=\linewidth]{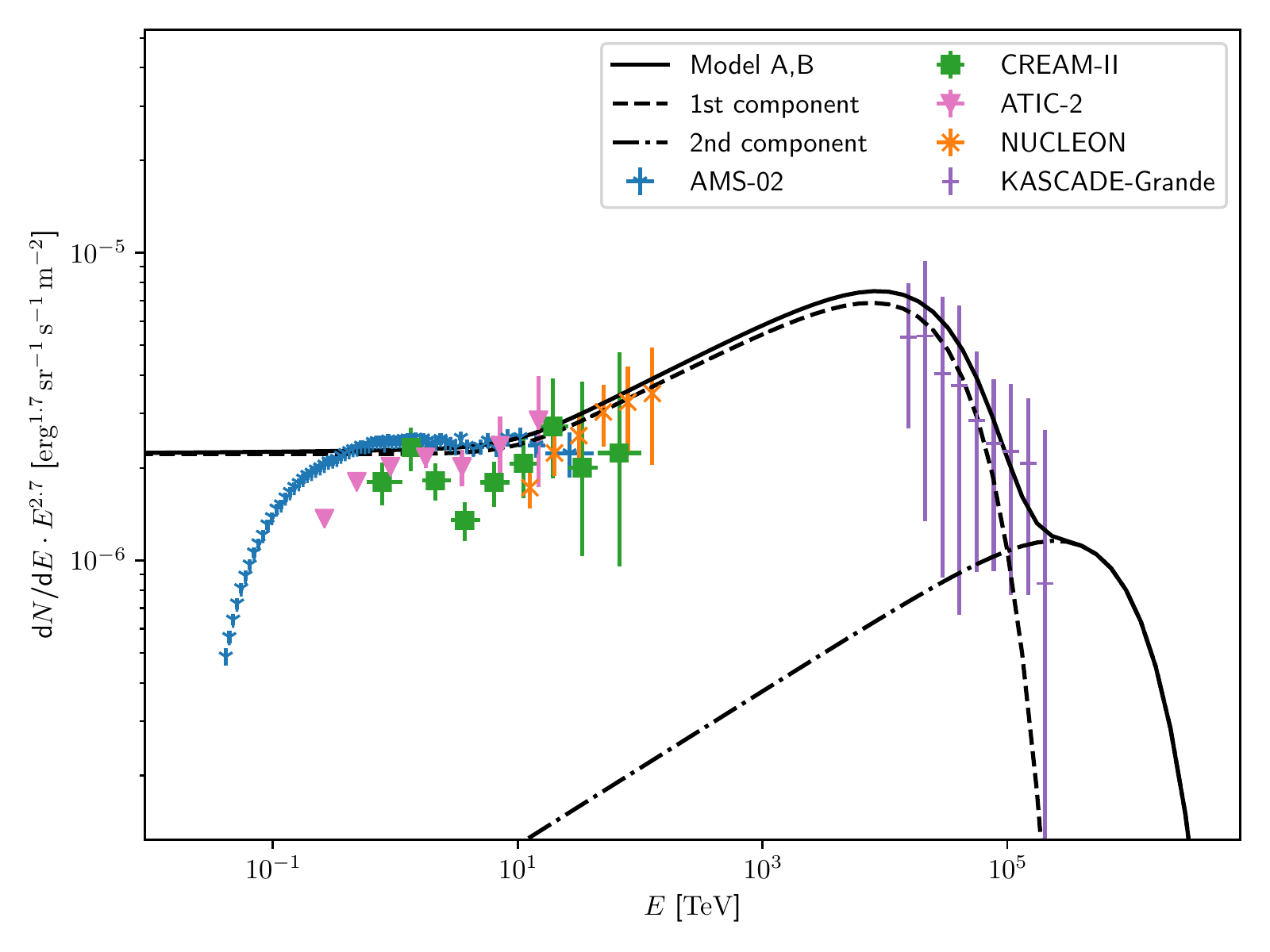}
    \caption{Silicon}
    \label{fig:CR_model_Si}
    \end{subfigure}
    \begin{subfigure}{0.49\textwidth}
    \centering
    \includegraphics[width=\linewidth]{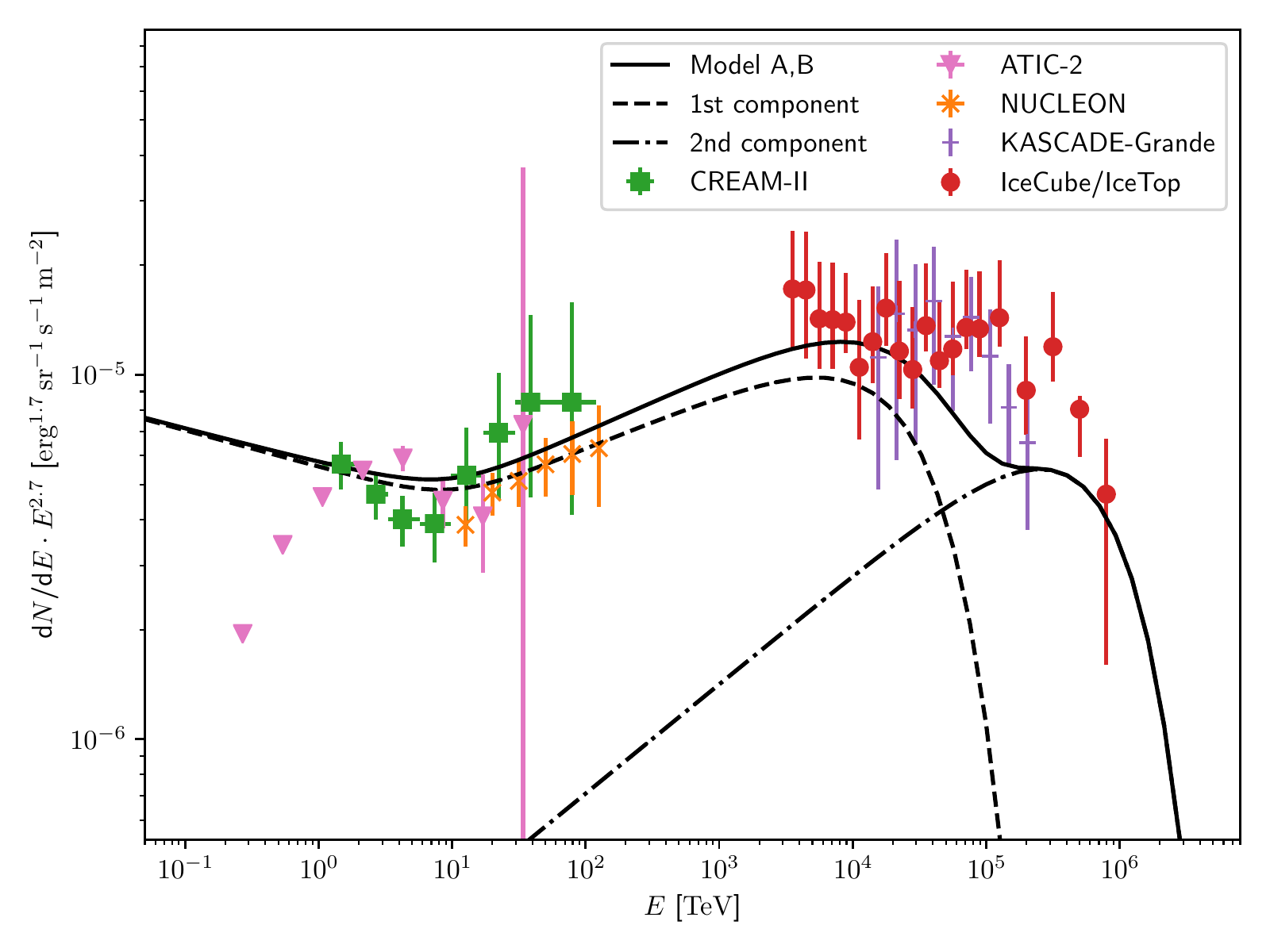}
    \caption{Iron}
    \label{fig:CR_model_Fe}
    \end{subfigure}
    \caption{Cosmic-ray model contributions from the heavier nuclei carbon, oxygen, magnesium, silicon, and iron together with the corresponding data from AMS-02 (\citealt{AMS02_2017} for C and O and \citealt{AMS02_2020} for Mg and Si), CREAM-II \citep{CREAM-II_2009}, ATIC-2 \citep{ATIC_2009}, NUCLEON \citep{Nucleon_data2019}, KASCADE-Grande \citep{KASCADE_data_2013}, and IceCube/IceTop \citep[for C-O and Fe]{IceTop_2019}. The contributions are the same in Model A and Model B. The first components are shown with the black dashed lines and the second components with the dashed-dotted lines, summing up to the total contributions (solid black lines).}
    \label{fig:model_components_heavier_nuclei}
\end{figure}

Figure \ref{fig:model_components_heavier_nuclei} shows the different contributions from the CR species C, O, Mg, Si, and Fe to the mixed CR Models A and B (described in the text and illustrated in Fig. \ref{fig:CR-model}) together with the CR data for each species. For IceCube/IceTop \citep{IceTop_2019} and KASCADE-Grande \citep{KASCADE_data_2013}, data are only given for individual elemental mass groups. We therefore assumed that data from the C-O group are split equally between C and O, and the same for Mg and Si and the corresponding Mg-Si group. The model parameters were not only adapted to fit the single-particle data, but also to match the all-particle data (see Fig. \ref{fig:CR-model}).

\FloatBarrier

\section{Emissivity of the CR models}
\begin{figure}[!h]
    \centering
    \includegraphics{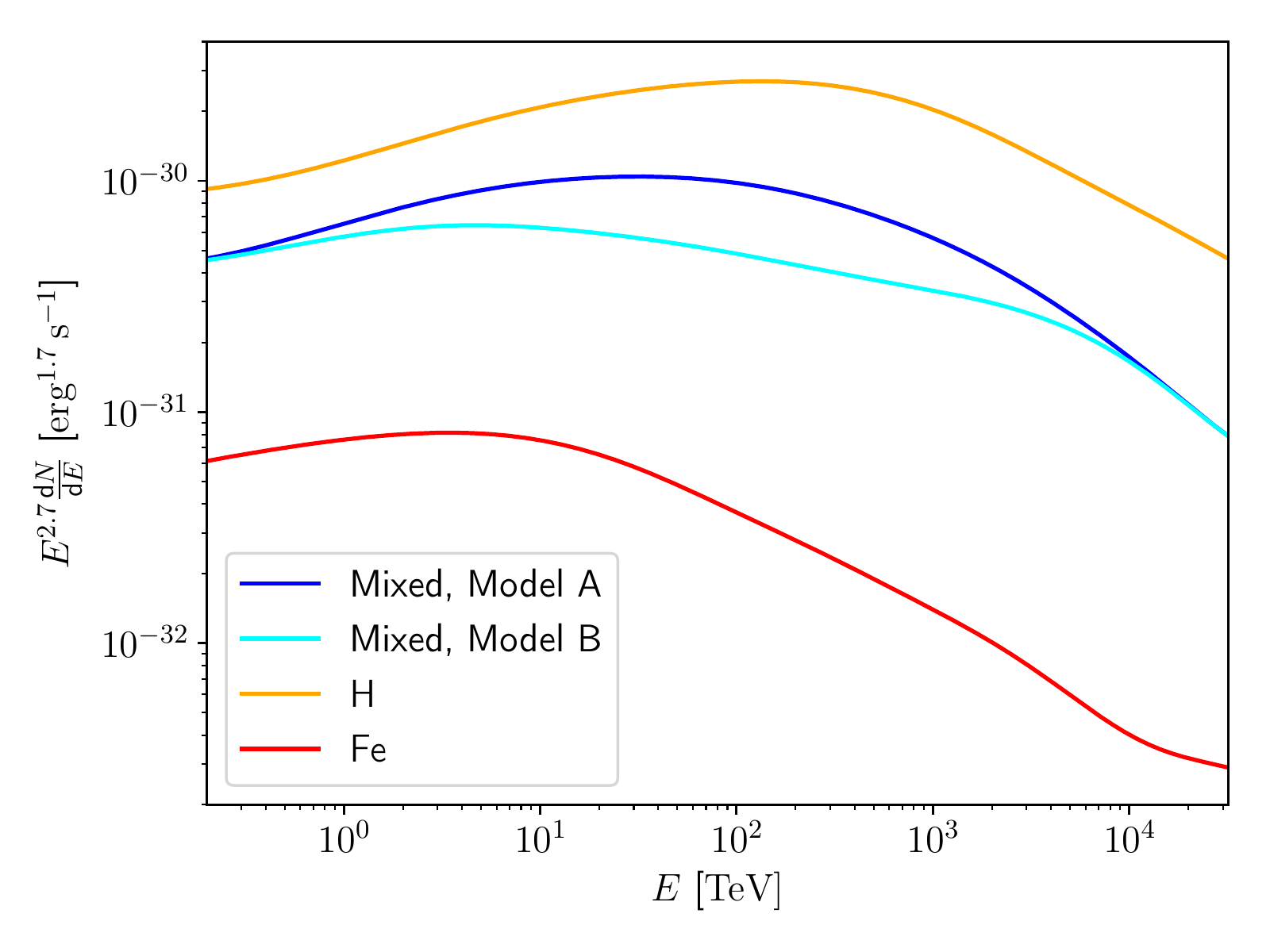}
    \caption{Gamma-ray production rates per \si{\cubic\centi\meter} per unit ISM number density from the mixed Model A (blue) and the mixed Model B (cyan) compared to the case assuming all CR particles from Model A are hydrogen (orange) or iron (red).}
    \label{fig:emissivity}
\end{figure}

In Fig. \ref{fig:emissivity} we show the total production rates per unity ISM density for $\gamma$ rays from the mixed CR Model A, the mixed CR Model B and by assuming all CRs from Model A are hydrogen and iron. The CR density is the local one, and the ISM composition is the one described in the text and used in Figs. \ref{fig:gamma-rays} and \ref{fig:neutrinos}. This shows the different emission levels, which are masked in Fig. \ref{fig:gamma-rays} due to the fit of the normalisation.

\FloatBarrier

\section{Impact of absorption for different CR distributions}\label{appendix_absorption}
\begin{figure}[!h]
    \centering
    \includegraphics[width=0.7\textwidth]{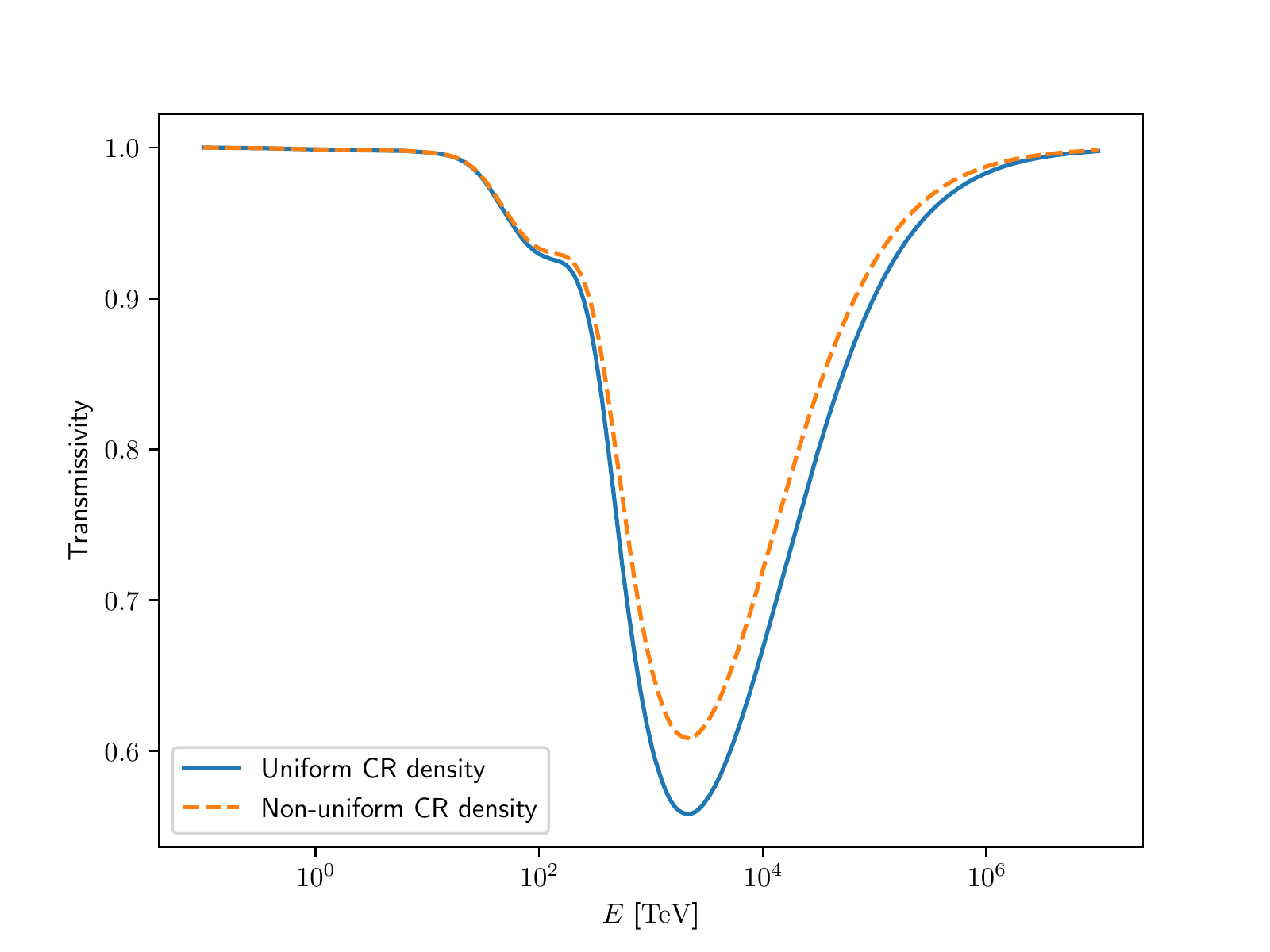}
    \caption{Comparison of the $\gamma$-ray transmissivities of the diffuse emission for a uniform CR density model and the distribution according to Eq. \ref{eq:CR_density}. The Galactic latitude and longitude range is the same as in Fig. \ref{fig:gamma-rays}, with Galactic longitudes $\SI{25}{\degree} \leq l \leq \SI{100}{\degree}$ and Galactic latitudes with $|b| \leq \SI{5}{\degree}$, and the ISM density model is the one from \cite{Ferriere1998} and \cite{Ferriere_et_al2007}.}
    \label{fig:transmissivities}
\end{figure}

The spatial distribution of CRs in the Milky Way influences the resulting absorption. To quantify these effects, we show in Fig. \ref{fig:transmissivities} the absorption for 
Galactic latitudes $|b| \leq \SI{5}{\degree}$, Galactic longitudes $\SI{25}{\degree} \leq l \leq \SI{100}{\degree}$, the ISM density model of \cite{Ferriere1998, Ferriere_et_al2007} and two different CR distributions: A uniform distribution, and a distribution satisfying Eq. \ref{eq:CR_density}, where the CRs are concentrated more towards the Galactic centre. The uniform CR distribution suffers slightly more absorption than the non-uniform distribution. This result might seem surprising at first glance, as the strong radiation fields in the Galactic centre result in $\gamma$ rays suffering strong absorption. However, $\gamma$ rays produced behind the Galactic centre are absorbed even more, because they not only have to traverse the stronger radiation fields at the centre, but also those behind the Galactic centre and the emission region. Furthermore, at \si{\peta\electronvolt} energies, the CMB dominates the absorption, and in this case the radiation field does not depend on the location. It is therefore the averaged and integrated local absorption, which determines the overall absorption. This can be understood within the framework of a slightly simplified calculation: We consider only absorption due to the CMB at \SI{1}{\peta\electronvolt} for the simplified case of a homogeneous ISM density. In this case, the total absorption is
\begin{align}
    e^{-\tau} = \frac{\int_{0}^{d_{\rm max}} \exp{(- \tau_{\rm CMB}\cdot x)} N_{\rm CR}(x) \text d x}{\int_{0}^{d_{\rm max}} N_{\rm CR}(x) \text d x}.
    \label{eq:absorption}
\end{align}
Here, $d_{\rm max}$ is the maximum distance from which Galactic $\gamma$-ray emission is received, $\tau_{\rm CMB} = \SI{3.575e-23}{\per\centi\meter}$ the optical depth of the CMB at \SI{1}{\peta\electronvolt,} and $N_{\rm CR}(x)$ the local CR density. For a uniform CR distribution, Eq. \ref{eq:absorption} simplifies to
\begin{align}
    e^{-\tau} 
    %= \frac{1}{d_{\rm max}}\int_0^{d_{\rm max}} \exp{(- \tau_{\rm CMB}\cdot x) \text d x}
    = \frac{1}{\tau_{\rm CMB}\cdot d_{\rm max}}\left[ 1 - \exp{(- \tau_{\rm CMB}\cdot d_{\rm max})}\right].
\end{align}
For the non-uniform CR density (Eq. \ref{eq:CR_density}, $z=0$), one obtains
\begin{align}
   & e^{-\tau} 
    %= \frac{\int_0^{d_{\rm max}} \exp{(- \tau_{\rm CMB}\cdot x) \sech{\frac{x-r_{\odot}}{R_{\rm CR}}} \text d x}}{\int_0^{d_{\rm max}} \sech{\frac{x-r_{\odot}}{R_{\rm CR}}} \text d x}
    = \frac{\left. \frac{-2 R_{\rm CR}}{R_{\rm CR}\tau_{\rm CMB}+1} \cdot \exp{((r_{\odot} - x(R_{\rm CR}\tau_{\rm CMB}+1))/R_{\rm CR}) \cdot {}_2F_1(1,a;b;c)} \right|_{x=0}^{d_{\rm max}}}
    {\left. R_{\rm CR} \arctan{\left(\sinh{\left( \frac{x-r_{\odot}}{R_{\rm CR}} \right)}   \right)}
    \right|_{x=0}^{d_{\rm max}}} \\
    &\text{with} \qquad a = \frac{1}{2}(R_{\rm CR}\tau_{\rm CMB}+1),
    \qquad b = \frac{1}{2}(R_{\rm CR}\tau_{\rm CMB}+3),
    \qquad c = - \exp{\left(\frac{2(r_{\odot}-x)}{R_{\rm CR}}\right)},
\end{align}
where ${}_2F_1$ is the Gaussian hypergeometric function. For $r_{\odot} = \SI{8}{kpc}$, $R_{\rm CR} = \SI{5.1}{kpc}$ and $d_{\rm max} = \SI{32}{kpc}$ this gives $e^{-\tau} = 0.275$ for the uniform CR distribution, and \num{0.407} for the non-uniform CRs. The uniform model suffers significantly more absorption, because a larger fraction of the $\gamma$-ray emission comes from larger distances behind the Galactic centre. Reducing $d_{\rm max}$ to \SI{16}{kpc}, which would mean that $\gamma$ rays are produced only within Galactic radii of \SI{8}{kpc}, leads to $e^{-\tau} = 0.470$ for the uniform CR distribution, and \num{0.457} for the non-uniform CRs. In this case, the uniform CR distribution suffers slightly less absorption. This illustrates the importance of the distances at which larger relative fractions of the $\gamma$-ray emission is produced. With realistic non-uniform density models and averaging over Galactic latitudes and longitudes, the differences in the absorption coefficients between different CR distributions are reduced. This is the case for the density model from \cite{Ferriere1998} and \cite{Ferriere_et_al2007} used in this work, where $\gamma$-ray production was considered out to Galactic radii of \SI{24}{kpc}, corresponding to $d_{\rm max} = \SI{32}{kpc}$ in the previous calculations.

\end{appendix}
\end{document}